\documentclass{JHEP3} 
\usepackage{epsfig}

\title {Contributions of the center vortices and vacuum domain in potentials between static sources }

\author{Seyed Mohsen Hosseini Nejad and Sedigheh Deldar \\ Department of Physics,\\ University of Tehran, P.O. Box 14395/547, Tehran 1439955961, Iran \\\\ E-mail: \email{smhosseininejad@ut.ac.ir, sdeldar@ut.ac.ir }  }

\abstract
{In this paper, we study the role of the domain structure of the Yang Mills vacuum. The Casimir scaling and $N$-ality are investigated in the potentials between static sources in various representations for $SU(2)$ and $SU(3)$ 
gauge groups based on the domain structure model using square ansatz for angle $\alpha_{C}(x)$. We also discuss about the contributions of the vacuum domain and center vortices in the static potentials. As a result, the potentials obtained from vacuum domains agree with Casimir scaling better than the ones obtained from center vortices. The reasons of these observations are investigated by studying the behavior of the potentials obtained from
vacuum domains and center vortices and the properties of the group factors. Then, the vacuum domains in $SU(N)$ and $G(2)$ gauge groups are compared and we
argue that the $G(2)$ vacuum is filled with center vortices of its subgroups.}

\preprint{}

\begin{document}
\section{Introduction}
 Mechanism of confinement is one of the challenging problems in QCD. Numerical methods (lattice gauge theory) and phenomenological models such as thick vortex 
model \cite{Fabe1998}, gluon chain model \cite{Greensite2002} and dual superconductor models (for a review, see ref. \cite{Ripka}) are investigating quark confinement. Popular 
degrees of freedom which are responsible for the confinement are center vortices, Abelian monopoles, instantons and merons. In this article, center vortices are studied as the 
confining degrees of freedom. 
 
There are numerical evidences in favor of the center vortex confinement mechanism (see refs. \cite{review} and \cite{michael} for review). The vortex model proposed by 't Hooft in the late $1970$'s \cite{Hoof79,Corn}, interprets confinement based on condensation 
of thin vortices with fluxes quantized in terms of the center elements of the gauge group (center vortices). A center vortex is
a color topological field configuration which is line like (in three dimensions) or surface like (in
four dimensions). Each thin center vortex piercing the Wilson loop leads to a certain amount of disorder and its effect on the loop is to multiply the loop
by a center element.
One of the criteria for the color confinement is the area law for the large Wilson loops:
\begin{equation}\label{wilson}
\langle W(C)\rangle~\sim~\exp{\big(-\sigma
Area(C)\big)}
\qquad \mbox{or} \qquad
V_{\bar{Q}Q}(R) \to~\sigma R\,.
\end{equation}
Here $C$ is a rectangular $R\times T$ loop in the $x-t$ plane,
$Area(C)$ is the minimal surface spanned on the loop $C$,
$\sigma> 0$ is the confining string tension, and $V_{\bar{Q}Q}(R)$ is the static
potential between sources at large distances $R$.

Using thin center vortex model, one gets an area law
for the potential of the quarks in the fundamental representation  but not for the adjoint representation.  According to the Monte Carlo data, confinement must be 
observed for color sources of higher representations \cite{Piccioni2005,Deldar1999,Bali2000}, as well. Modifying the model by applying finite thicknesses
to the thin vortices,
the area law was observed for the Wilson loops of all representations \cite{Fabe1998}. 

Furthermore, by adding vacuum domains corresponding to the trivial center element $z_0=1$, confinement interval has been increased  \cite{Greensite2007}.  
Vacuum domains carry quantized magnetic fluxes in terms of the trivial center element. This modified model has been called  domain structure model in the literature.  
We study the characteristics of the vacuum domain and their effects on the potentials between color sources. $G(2)$ 
gauge group is its own universal covering group and it has only one trivial center element. Therefore $G(2)$ is 
an interesting laboratory, which attracts considerable attentions to 
examine the role of the trivial center element \cite{Greensite2007,Holland2003jy,Pepe2005sz,Pepe2006er,Cossu2006bi,Cossu2007dk,Cossu2007ns,Maas2007af}. One  expects that in $G(2)$ gauge group the static potentials do not grow
linearly over any certain range of distances because of the lack of nontrivial center element. However, the linear rise of the  potentials for all representations at the
intermediate distances was clearly observed in lattice gauge theory \cite{Greensite2007,Olejnik2008}. On the other hand, from the domain structure model, we have observed linear 
potentials between $G(2)$ sources for different representations roughly proportional to the eigenvalue of the quadratic Casimir operator
of the representation which is in agreement with the lattice results \cite{Deldar2012}. In our previous article \cite{Deldar2014}, We have argued that the $SU(2)$ and $SU(3)$
subgroups of the $G(2)$ gauge group may be responsible for the
confinement potential of $G(2)$. We applied the domain structure model to the $G(2)$ gauge group and the thick vortex model to the $SU(2)$ and
$SU(3)$ subgroups of $G(2)$. We discussed about the reasons of observing linear potential in $G(2)$ gauge group by comparing the potentials and extremums of the vortex profile ${\mathrm {Re}}(g_{r})$ of
the $G(2)$ gauge group and its subgroups in the fundamental ($\{7\}$-dimensional) and adjoint ($\{14\}$-dimensional) representations. 

The vacuum domain may play an important role not only in $G(2)$ but also in $SU(N)$ gauge
theories. In this paper, we study the role of the vacuum domain at intermediate regime for $SU(N)$ gauge groups and we discuss about the possibility of constructing a vacuum domain by center 
vortices. Then, the vacuum domain of $G(2)$ is studied. By comparing with  $SU(2)$ and  $SU(3)$ groups, the role of these subgroups in observing confinement in $G(2)$ is discussed.  In Sect.~\ref{2}, 
we briefly review  the domain structure model. Angle parameter of the model is studied in Sect.~\ref{3}. Then, in 
Sect.~\ref{4}, we obtain static potentials for various representations and their ratios for different kind of  center domains for the $SU(2)$ and $SU(3)$ gauge groups. 
The interaction between the Wilson loops and vacuum domains and a comparison between $SU(N)$ and $G(2)$ groups are discussed in Sect.~\ref{5}. We summarize the main points of 
our study in Sect.~\ref{6}. Finally,  Cartan generators are constructed using tensor product  and decomposition methods in the appendix. 
\section{ Domain structure model of the Yang Mills vacuum }
\label{2}
In this model, the vacuum is assumed to be filled with domain structures. In $SU(N)$ gauge group, there are $N$ types of center domains including center vortices  corresponding to 
the nontrivial center elements of $\mathbb Z_N$ subgroup enumerated by the value $n=1,...,N-1$ and the vacuum type corresponding to
the $z_0 = 1$ center element ($n=0$). For $G(2)$ gauge group, there is of course only one center domain of vacuum type corresponding to 
$z_0= 1$ which belongs to the trivial $\mathbb Z_1$ subgroup. The probability that any given plaquette is pierced by an
$n$th domain is equal to $f_n$. Creation of a thick center domain linked to
a Wilson loop in representation $r$ has the 
effect of multiplying the Wilson loop by a group factor $\mathcal{G}_r(\alpha^{(n)})$, $\it{i.e.}$ 

\begin{equation}
\label{group factor}
W_r(C) \to~ \mathcal{G}_r(\alpha^{(n)}) W_r(C)= \frac{1}{d_r}\mathrm{Tr}\left(\exp\left[i\vec{\alpha}^{(n)}\cdot\vec{\mathcal{H}}\right]\right) W_r(C),
\end{equation}
where the $\{\mathcal{H}_i~i=1,..,N-1\}$ are the Cartan generators,
 angle $\vec{\alpha}^{(n)}$ shows the flux profile that depends on the location of the $n$th center domain with respect to the Wilson loop, and $d_r$ is the dimension of the representation 
$r$. If the core of the center domain is entirely enclosed by the loop, then 
\begin{equation}
\mathcal{G}_r(\alpha^{(n)}) =(z_n)^k=e^{\frac{i2k\pi n}{N}},
\end{equation}
 where $k$ is the $N$-ality of the representation $r$ and if the core is entirely outside
  the minimal area of the loop, then the group factor is equal to 1. 

Phase factors of domains of type $n$ and type $N-n$ are complex conjugates of each other and they may be considered
as the same type of domains but with magnetic flux pointing in opposite directions, 
so that
\begin{equation}
\label{equal}
       f_n = f_{N-n}  ~~~~ \mbox{and} ~~~~ 
       \mathcal{G}_r[\alpha_C^n(x)] = \mathcal{G}_r^*[\alpha_C^{N-n}(x)].
\end{equation}
The inter quark potential induced by the center domains is as the following \cite{Fabe1998,Greensite2007}:
\begin{equation}
\label{potential}
V_r(R) = -\sum_{x}\ln\left\{ 1 - \sum^{N-1}_{n=0} f_{n}
(1 - {\mathrm {Re}}\mathcal{G}_{r} [\vec{\alpha}^n_{C}(x)])\right\},
\end{equation}
where the function $\vec{\alpha}^n_{C}(x)$ represents the corresponding angle and
depends on both  the Wilson contour $C$ and  the position of the vortex center $x$. In the next section, some reasonable
ansatz for the angle $\vec{\alpha}^n_{C}(x)$ are given.
\section{ Ansatz for the angle $\alpha_{C}(x)$ }
\label{3}
\begin{figure}
\begin{center}
\resizebox{0.6\textwidth}{!}{
\includegraphics{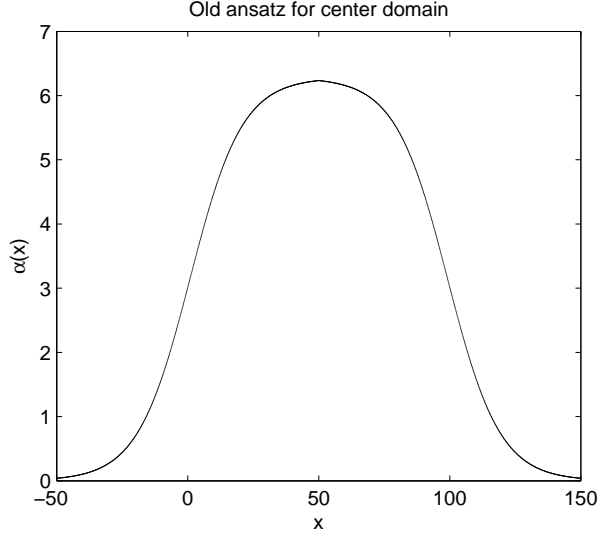}}
\caption{
The angle $\alpha(x)$ versus $x$ obtained from the old ansatz. The free parameters are $R=100$, $a=0.05$, and $b=4$.}
\label{C1}
\end{center}
\end{figure}

There are some functions to use as the ansatz for $\alpha_{C}(x)$ \cite{Fabe1998,Greensite2007,Deldar2001}. An appropriate ansatz must lead to a well-defined potential $\it{i.e.}$ respecting linearity 
and Casimir scaling for the intermediate regime. The Wilson contour $C$ is a rectangular $R\times T$ with $T>>R$ and left and right timelike legs of the
 loop are located at $x=0$ and $x=R$, respectively. A few conditions that any ansatz must satisfy are as the followings: 
\begin{description}
\item{1.} If a center domain locates outside the minimal area of the Wilson loop, then $\alpha_{C}(x)= 0$. 

\item{2.} If the minimal area of the Wilson loop is pierced by a center domain, then $\vec\alpha_{C}(x)=\vec\alpha_{max}$, where $\vec\alpha_{max}$ is
obtained from the following maximum flux condition:
\begin{equation}
\exp(i\vec{\alpha}_{max}\cdot\vec{H_r})=e^{i2k\pi n /N} I.
\end{equation}
\item{3.} If $R\to0$, then the flux of the domain core must be zero inside the Wilson loop $\it{i.e.}$ $\alpha_{c}(x) \to~0$.
\end{description}
An ansatz introduced by Faber $\it{et~ al.}$ \cite{Fabe1998} is:

\begin{equation}
\label{Oansax}
\alpha^i_R(x) = {{\alpha}^i_{max}\over 2} \left[1 - \tanh\left(ay(x) + {b\over R}\right) \right],
\end{equation}
where $a$ and $b$ are free parameters and $y(x)$ is 

\begin{equation}
  y(x) = \left\{ \begin{array}{cl}
                     x-R  & \mbox{for~} |R-x| \le |x| \cr
                      -x & \mbox{for~} |R-x| > |x|
                   \end{array} \right\}. 
\end{equation}
The magnitude of $y(x)$  shows the distance of the center of domain with respect to the nearest
timelike leg of the Wilson loop. Figure \ref{C1} shows this old ansatz versus $x$ for $R=100$.

\begin{figure}
\begin{center}
\resizebox{0.65\textwidth}{!}{
\includegraphics{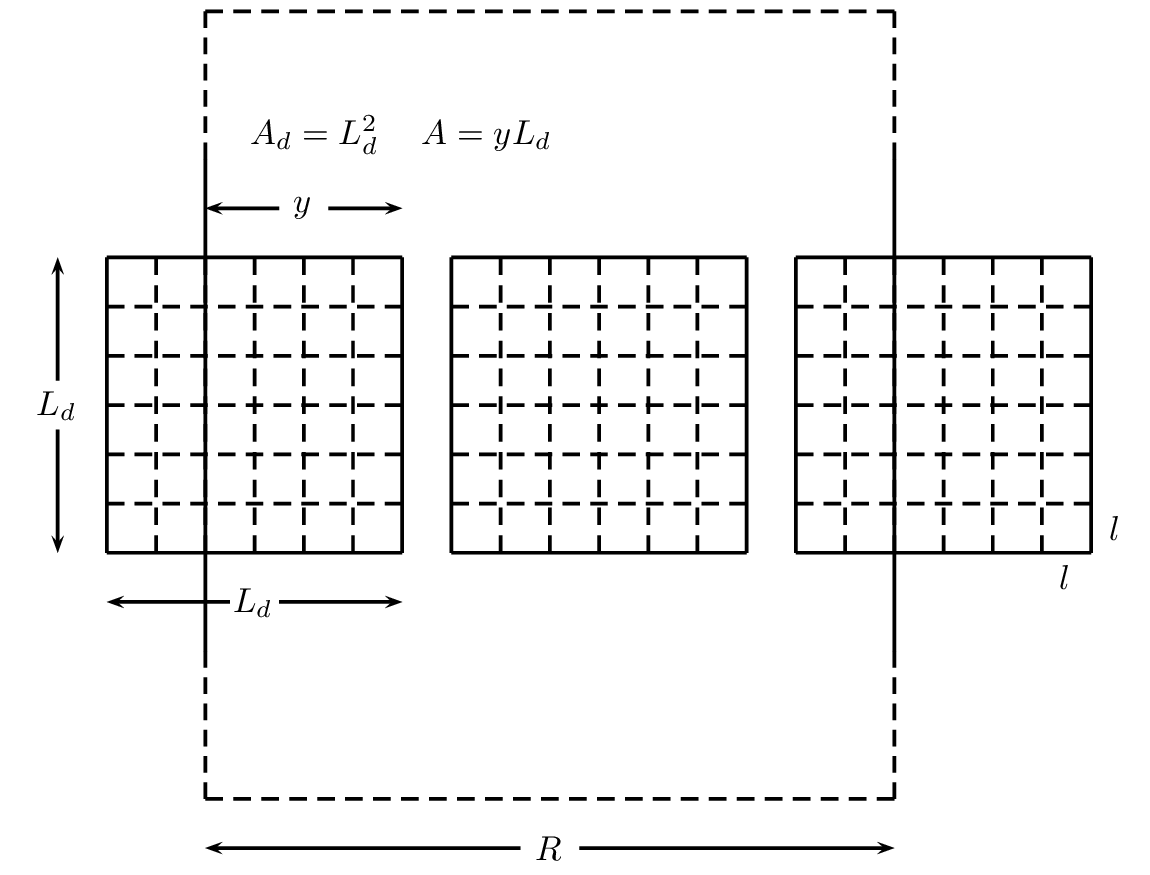}}
\caption{\label{FIG1}
The figure schematically shows the interaction between the angle of square ansatz and the Wilson loop as well as some parameters of square ansatz. }
\end{center}
\end{figure}   
Another ansatz was introduced by Greensite $\it{et~ al.}$ \cite{Greensite2007}. Each
domain, with cross section $A_d$, is divided to subregions of area $l^2\ll A_d$ which $l$ is a short correlation length. The color magnetic fluxes in subregions $l^2$ fluctuate randomly and 
almost independently. In other word, the color magnetic fluxes in neighboring regions of area $l^2$ are uncorrelated. The only constraint is that the total color magnetic fluxes of the subregions 
must correspond to an element of the gauge group center. The ansatz is introduced as the following:  
\begin{equation}
\label{Sansax}
         \vec{\alpha}^n_C(x)\cdot\vec{\alpha}^n_C(x) = \frac{A_d}{ 2\mu} \left[
\frac{A}{ A_d} - \frac{A^2}{ A_d^2} \right]
                         + \left(\alpha^n_{max} \frac{A}{ A_d}\right)^2,
\end{equation}
where $A$ is the cross section of the center domain overlapping with the minimal area of the Wilson loop and $\mu$ is a free parameter. The cross section 
of a domain is  a $L_d \times L_d$ square. Figure \ref{FIG1}
schematically shows the interaction between the angle of square ansatz and the Wilson loop. Now one should take two intervals for the square ansatz: 
\begin{figure*}[t]
\epsfxsize=8cm 
\epsffile{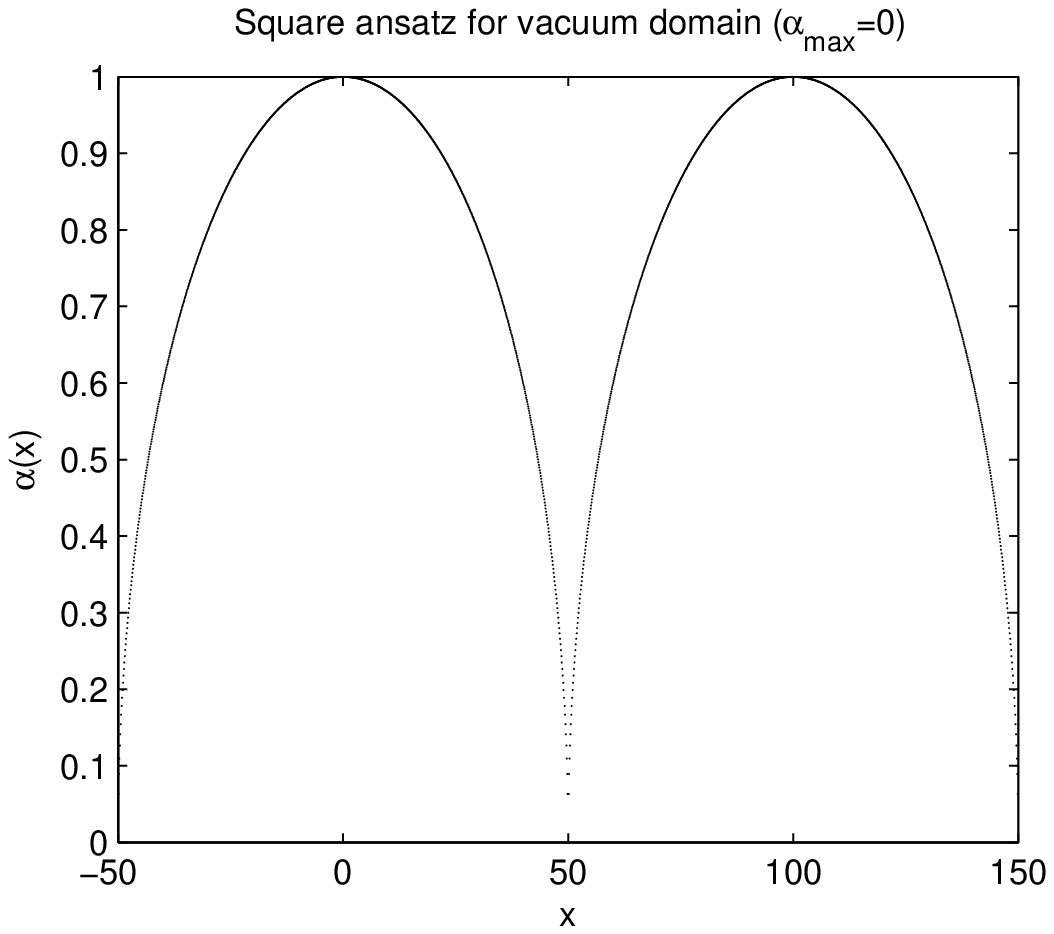} \hspace{0.0cm}
\epsfxsize=8cm 
\epsffile{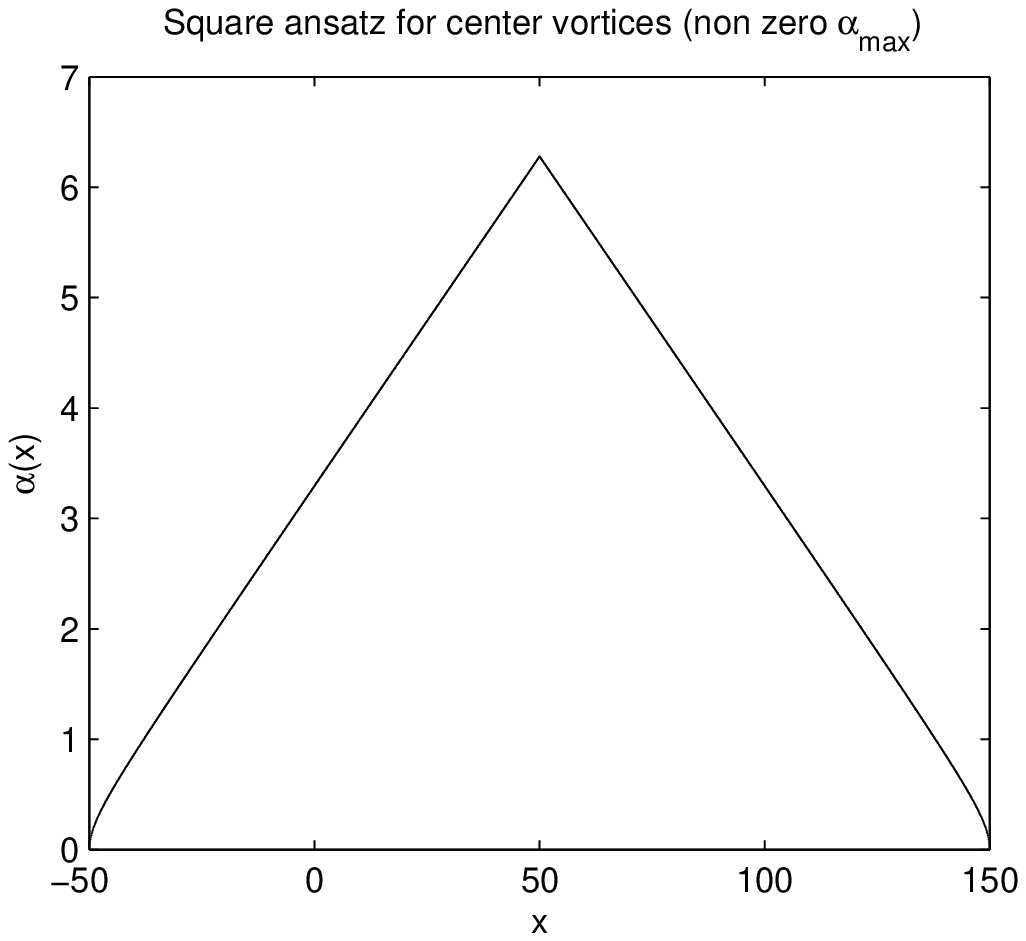} 
\caption{ Angle $\alpha(x)$  versus $x$. Left panel represents the square ansatz for $\alpha_{max}=0$ (vacuum domain). Right panel represents the square ansatz for non zero $\alpha_{max}$ (center vortices). The free parameters are $R=100$, $L_{d}=100$, and $L^{2}_{d}/(2\mu)=4$.}
\label{A3}
\end{figure*}
\begin{itemize}   
\item $\mbox{for~} 0 \le R\le L_d$,
\begin{equation}
  \vec{\alpha}^n_C(x)\cdot\vec{\alpha}^n_C(x) = \left\{ \begin{array}{cl}
                     \frac{{L_d}^2}{ 2\mu} \left[
\frac{y{L_d}}{ {L_d}^2} - \frac{({y{L_d})}^2}{ {L_d}^4} \right]
                         + \left(\alpha^n_{max} \frac{{y{L_d}}}{ {L_d}^2}\right)^2 & \mbox{for~} -\frac{{L_d}}{2} \le x\le -\frac{{L_d}}{2}+R \cr
                      \frac{{L_d}^2}{ 2\mu} \left[
\frac{RL_d}{ {L_d}^2} - \frac{{RL_d}^2}{ {L_d}^4} \right]
                         + \left(\alpha^n_{max} \frac{{RL_d}}{ {L_d}^2}\right)^2  & \mbox{for~} -\frac{{L_d}}{2}+R \le x\le \frac{{L_d}}{2}\cr
                         \frac{{L_d}^2}{ 2\mu} \left[
\frac{y{L_d}}{ {L_d}^2} - \frac{({y{L_d})}^2}{ {L_d}^4} \right]
                         + \left(\alpha^n_{max} \frac{{y{L_d}}}{ {L_d}^2}\right)^2 & \mbox{for~} \frac{{L_d}}{2} \le x\le R+\frac{{L_d}}{2}
                   \end{array} \right\} ,
\end{equation}
\item $\mbox{for~} L_d\le R $
\begin{equation}
  \vec{\alpha}^n_C(x)\cdot\vec{\alpha}^n_C(x) = \left\{ \begin{array}{cl}
                      \frac{{L_d}^2}{ 2\mu} \left[
\frac{y{L_d}}{ {L_d}^2} - \frac{({y{L_d})}^2}{ {L_d}^4} \right]
                         + \left(\alpha^n_{max} \frac{{y{L_d}}}{ {L_d}^2}\right)^2 & \mbox{for~} -\frac{{L_d}}{2} \le x\le \frac{{L_d}}{2} \cr
                       (\alpha^n_{max})^2  & \mbox{for~} \frac{{L_d}}{2} \le x\le R-\frac{{L_d}}{2}\cr
                         \frac{{L_d}^2}{ 2\mu} \left[
\frac{y{L_d}}{ {L_d}^2} - \frac{({y{L_d})}^2}{ {L_d}^4} \right]
                         + \left(\alpha^n_{max} \frac{{y{L_d}}}{ {L_d}^2}\right)^2 & \mbox{for~} R-\frac{{L_d}}{2} \le x\le R+\frac{{L_d}}{2}
                 \end{array} \right\} ,
\end{equation}

\end{itemize}
where
\begin{equation}
  y(x) = \left\{ \begin{array}{cl}
                    R-x+\frac{{L_d}}{2}  & \mbox{for~} |R-x| \le |x| \cr
                      x+\frac{{L_d}}{2} & \mbox{for~} |R-x| > |x|
                   \end{array} \right\}.
\end{equation}
The range of $x$ $\it{i.e.}$ $-\frac{{L_d}}{2} \le x\le R+\frac{{L_d}}{2}$ has been restricted over all plaquettes within the minimal area of the Wilson loop, as well as plaquettes
in the plane outside the perimeter of the loop which are located inside a distance $\frac{{L_d}}{2}$ of the loop. Figure \ref{A3} shows this square ansatz versus $x$ for $R=200$. Center domains are located completely inside the Wilson loop at $x=0$. The angle $\alpha(x)$ changes more drastically in the right plot where it is obtained by the center vortices (non zero $\alpha_{max}$) compared with the left plot where the vacuum domains ($\alpha_{max}=0$) are used. In the next 
section, we argue about the contribution of the vacuum domain to the potential between color sources at intermediate distances for the $SU(N)$ gauge theories.

\section{ Static potentials and Casimir scaling } 
\label{4}
 The center vortex model \cite{Fabe1998} leads to linear regime for the static potential qualitatively in agreement with Casimir scaling 
hypothesis. The confinement regime has been increased \cite{Greensite2007} when the vacuum domains have been added to the model. In our 
previous papers \cite{Deldar2012,Deldar2014}, we have studied the role of the vacuum domain in $G(2)$ gauge group which has one trivial center element, only. According to the center vortex theory, one does not expect confinement in a group
without nontrivial center element. But using the domain model and from the numerical lattice calculations for the $G(2)$ gauge group, the static potentials in different representations 
grow linearly at intermediate distances and the ratios of the linear regime slopes are roughly proportional to the Casimir ratios. Therefore, it is interesting to understand the
 role of the vacuum domain to the static potential in $SU(N)$ gauge theories. If one uses the square ansatz $\it{i.e.}$ Eq. (\ref {Sansax}), then the static potential induced by center 
vortices is as the following \cite{Fabe1998}:     
\begin{equation}
\label{potential 2}
V_r(R) =  - \sum^{{ {L_d}/2 + R}}_{{x=-{L_d}/2}} \ln\left\{ 1 - \sum^{N-1}_{n=1} f_{n}
(1 - {\mathrm {Re}} \mathcal{G}_{r} [\vec{\alpha}^n_{C}(x)])\right\},
\end{equation}
and the contribution of the vacuum domain added to the static potential is given by \cite{Greensite2007}:
\begin{equation}
         V_r(R) =  - \sum^{{ {L_d}/2 + R}}_{{x=-{L_d}/2}} \ln\Bigl\{(1-f_0) + f_0{\mathrm {Re}}\mathcal{G}_{r}[\vec\alpha_C^0(x)]\Bigr\}, 
\label{Vr}
\end{equation}
where $f_0$ is the probability that any given unit is pierced by a vacuum domain. Now, we obtain the static potential 
at different distances in $SU(N)$ gauge group $(N=2,3)$, using the contributions of  all domains, vacuum domain and center vortices, separately.

\subsection{$SU(2)$ case}
\label{sub1}
First, we apply the model to the $SU(2)$ gauge group. In $SU(2)$ case, there is one nontrivial
center element in addition to the trivial  element
\begin{equation}
\mathbb Z(2)=\{z_0=1,z_1=e^{\pi i}\}.
\end{equation}
Therefore, the static potential induced by all domains of $SU(2)$ gauge group is obtained from Eq. (\ref {potential}) 
\begin{equation}
V_j(R) =  - \sum^{{ {L_d}/2 + R}}_{{x=-{L_d}/2}} \ln\Bigl\{(1-f_0-f_1) + f_0{\mathrm {Re}}\mathcal{G}_j[\alpha^{0}_C(x)] 
+ f_1{\mathrm {Re}}\mathcal{G}_j[\alpha^{1}_C(x)]\Bigr\}, 
\end{equation}
where $f_1$ and $f_0$ are the
probabilities that any given unit area is pierced by a center vortex and a vacuum domain, respectively. The free parameters $L_{d}$, $f_{1}$, $f_{0}$, and $L^{2}_{d}/(2\mu)$ 
 are chosen to be $100$, $0.01$, $0.03$, and $4$, respectively. We take the  correlation length $l=1$, therefore the static potentials are linear from the
beginning ($R=l$). The square ansatz for the angles corresponding to the Cartan generator $\mathcal{H}_3$ for the center vortex and the vacuum domain are:
\begin{equation}
         \Bigl((\alpha^1_C(x)\Bigr)^{2} = \frac{A_d}{ 2\mu} \left[
\frac{A}{ A_d} - \frac{A^2}{ A_d^2} \right]
                         + \left(2\pi \frac{A}{ A_d}\right)^2,~~~
 \\
         \Bigl((\alpha^0_C(x)\Bigr)^{2} = \frac{A_d}{ 2\mu} \left[
\frac{A}{ A_d} - \frac{A^2}{ A^{2}_d} \right].                          
\end{equation}

\begin{figure}
\begin{center}
\resizebox{0.6\textwidth}{!}{
\includegraphics{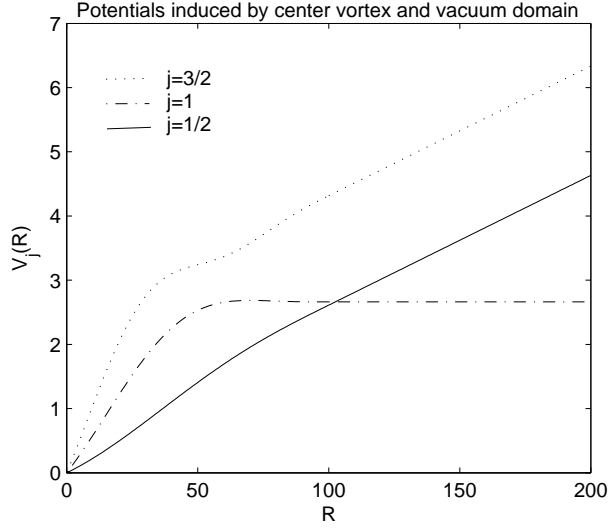}}
\caption{\label{FIG2}
The static potential between static sources induced by all domains (center vortex and vacuum domain) for some representations of $SU(2)$ gauge group. The free parameters are 
$L_{d}=100$, $f_{0}=0.03$, $f_{1}=0.01$ and $L^{2}_{d}/(2\mu)=4$ \cite{Greensite2007}.  
}
\end{center}
\end{figure}

\begin{figure}
\begin{center}
\resizebox{0.6\textwidth}{!}{
\includegraphics{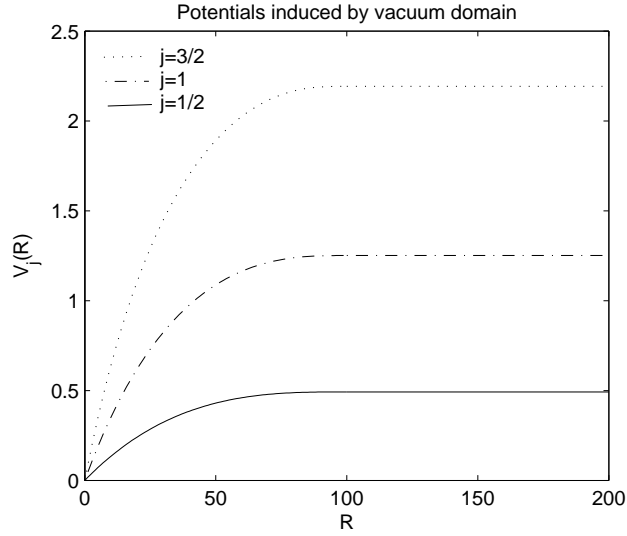}}
\caption{\label{FIG3}
The static potential induced by the vacuum domain. The potentials are screened
at large distances where the vacuum domain is located
completely inside the Wilson loop. At intermediate distances, $R \in [0,20]$, where the vacuum domain is partially located
inside the Wilson loop, a linear regime is observed. The free parameters are chosen to be $L_{d}=100$, $f_{0}=0.03$, and $L^{2}_{d}/(2\mu)=4$.
}
\end{center}
\end{figure}

\begin{figure*}[t]
\epsfxsize=8cm 
\epsffile{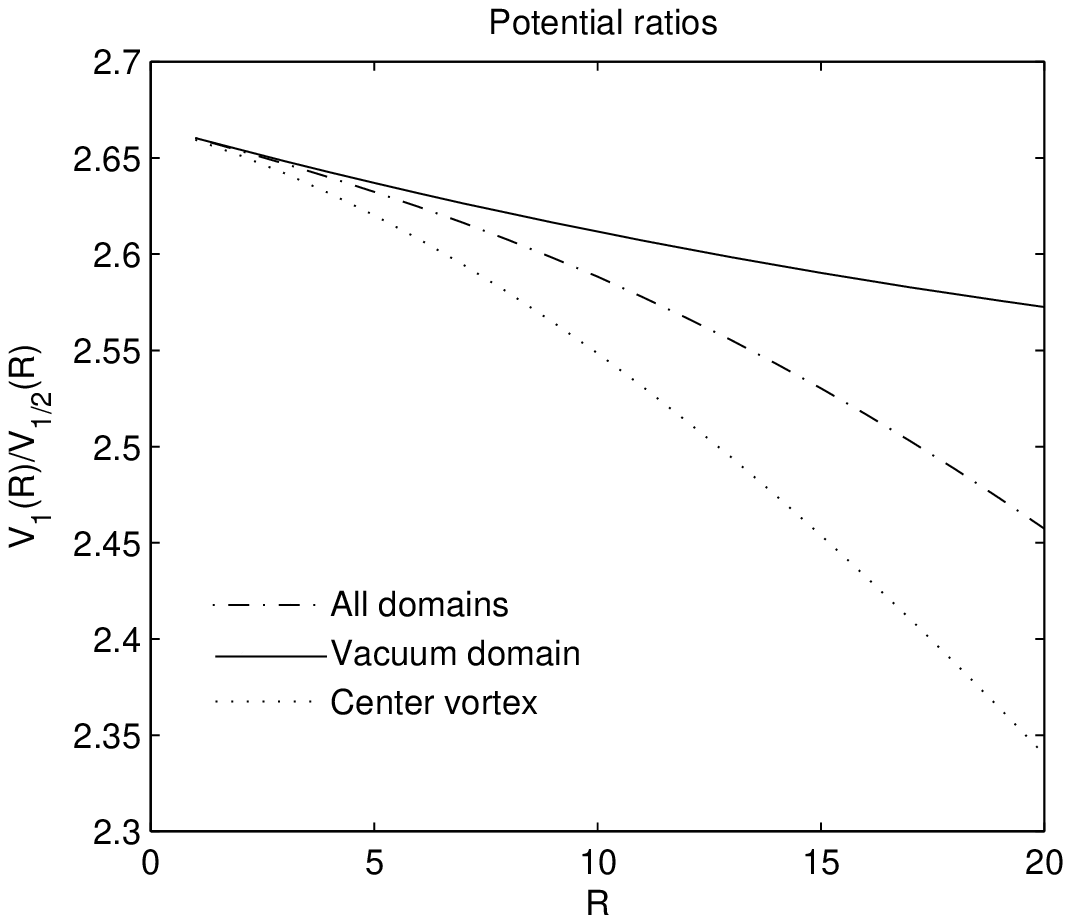} \hspace{0.0cm}
\epsfxsize=8cm 
\epsffile{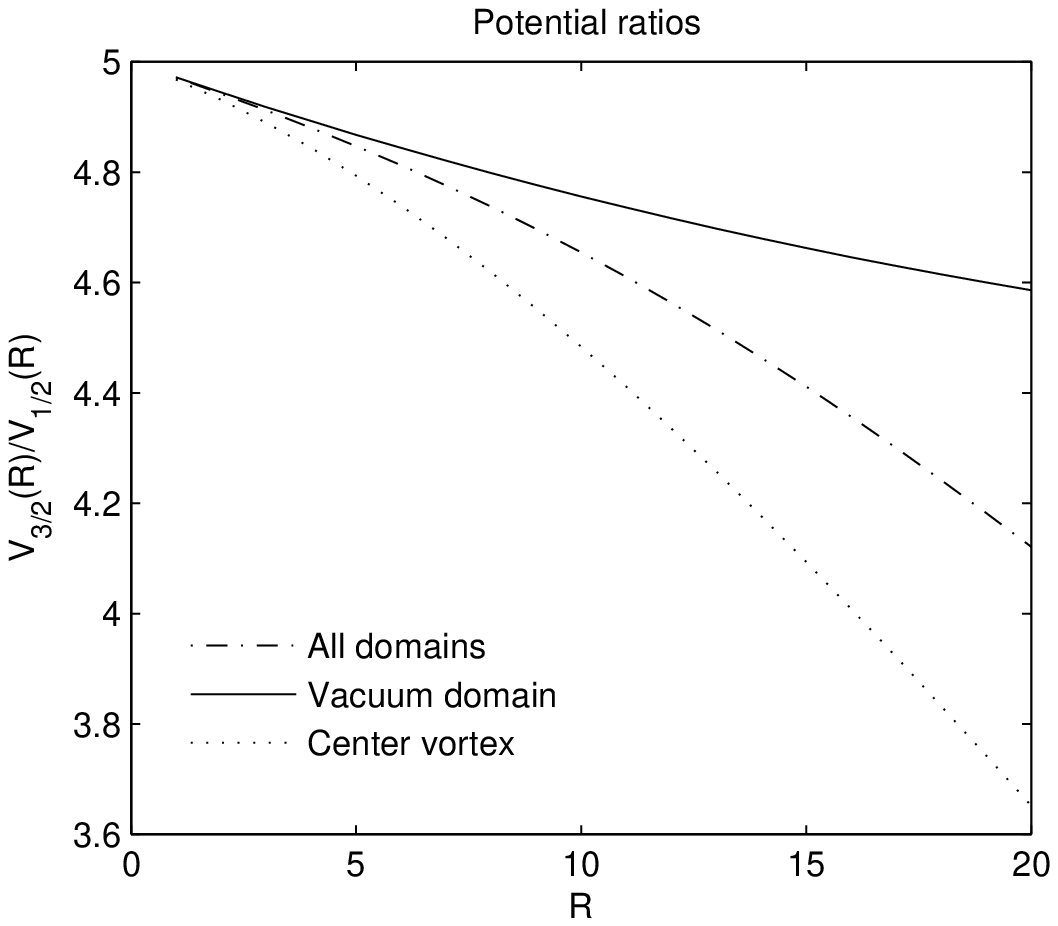} 
\caption{Left panel: ratios of ${V_1(R)}/{V_{1/2}(R)}$ are shown. Upper curve shows the contribution of the vacuum domain which starts from the Casimir ratios ($8/3$) and violates very slowly from the Casimir 
ratio compared with the potentials obtained from all domains (middle curve) and the center vortex (lower curve). Right panel: ratios of ${V_{3/2}(R)}/{V_{1/2}(R)}$  
are shown. The same arguments are true for the $j=3/2$ representation. The free parameters are $L_{d}=100$, $f_{0}=0.03$, $f_{1}=0.01$ and $L^{2}_{d}/(2\mu)=4$.}
\label{Y1}
\end{figure*}

\begin{figure}
\begin{center}
\resizebox{0.6\textwidth}{!}{
\includegraphics{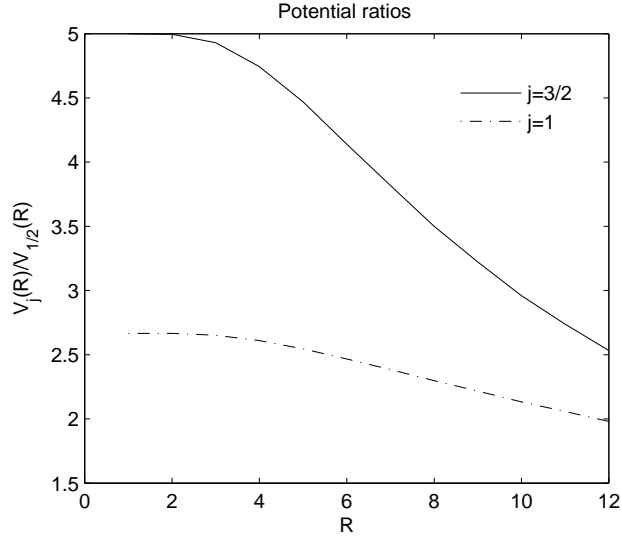}}
\caption{\label{D1}
Ratios of ${V_j(R)}/{V_{1/2}(R)}$ induced by center vortex for adjoint and $j=3/2$ representations using  old ansatz for angle $\alpha_{C}(x)$. The selected free parameters are $f_1=0.1,~a=0.05,~b=4$ \cite{Fabe1998}. The potential ratios start from the Casimir ratios but the ratios by this ansatz drop steeper than the ratios by square ansatz. 
}
\end{center}
\end{figure}

Figures \ref{FIG2} and \ref{FIG3} show the static potentials $V_{j}(R)$ for the $j = 1/2, 1, 3/2$ representations using all domains and vacuum domain, 
respectively. At intermediate distances, the potentials induced by all domains are linear in the range $R\in [0,20]$ \cite{Greensite2007}. For the same interval, the potentials induced by 
vacuum domain are also linear. The potentials are screened
at large distances where the vacuum domain is located
completely inside the Wilson loop. At intermediate distances, $R \in [0,20]$, where the vacuum domain is partially located
inside the Wilson loop, a linear regime is observed. Figure \ref{Y1} plots the ratios $V_1(R)/V_{1/2}(R)$ (left panel) and $V_{3/2}(R)/V_{1/2}(R)$ (right panel) for the linear regime using
vacuum domain, center vortex, and all domains. 
These potential ratios start from the ratios
of the corresponding Casimirs i.e. 
\begin{equation}
\frac{C_{1}}{C_{1/2}}=8/3, ~~~~~~~~~~~~~~~~~~~~~~~\frac{C_{3/2}}{C_{1/2}}=5.
\end{equation}
In the range $R\in [0,20]$, the potential ratios $V_1(R)/V_{1/2}(R)$ and $V_{3/2}(R)/V_{1/2}(R)$ induced by center vortices decrease slowly from $8/3$ and $5$ to about $2.34$ and $3.65$, respectively. 
In the same interval, the potential ratios $V_1(R)/V_{1/2}(R)$ and $V_{3/2}(R)/V_{1/2}(R)$ induced by vacuum domain drop very slowly from $8/3$ and $5$ to about $2.57$ and $4.6$ compared with the potential ratios induced by center vortices. On the other hand, Fig. \ref{D1} shows potential ratios using the ansatz given in Eq. (\ref {Oansax}), for the choice of parameters 
$f=0.1$,~$a=0.05$, and~$b=4$. The potential ratios $V_1(R)/V_{1/2}(R)$ and $V_{3/2}(R)/V_{1/2}(R)$ induced by center vortices drop from $8/3$ and $5$ to about $2$ and $2.5$ in the range $R \in [1,12]$, respectively 
\cite{Fabe1998}. So the potential ratios obtained from square ansatz drop slower than the ones by the old ansatz.  

From Fig. \ref{FIG2}, it is clear that at large distances, $R\geq 100$, the static potentials induced by all domains agree with $N$-ality as expected. Therefore, the main contribution to the potentials for large loops corresponds to center vortices. 
$N$-ality classifies the representations of a gauge group. At large distances, when the energy between two static sources is equal or greater than twice the 
gluon mass, a pair of gluon-anti gluon are popped out of the vacuum and combine with initial sources and transform them into the lowest order representations of their class. 
For examples
\begin{equation}
\{3\} \otimes \{3\}=\{1\} \oplus \{3\} \oplus \{5\},
\label{adjoint3}
\end{equation}
\begin{equation}
\{4\} \otimes \{3\}=\{2\} \oplus \{4\} \oplus \{6\}.
\label{32}
\end{equation}
In other words, static sources in representations $\{4\} (j=3/2)$ and $\{3\} (j=1)$ by combining with a gluon are transformed into the lowest order
representation $\{2\} (j=1/2)$ and color singlet. Thus, the slope of representation $\{4\}$ must be the same as the fundamental one and representation $\{3\}$ 
must be screened. Screening is observed in Fig. \ref{FIG3}, since vacuum domain 
locates completely inside the Wilson loop at large distances.

Therefore for SU($2$) case, the fluctuations within a vacuum domain lead to a group disorder which agrees Casimir scaling stronger than center vortices while center vortex disorder leads to 
$N$-ality. 

\subsection{ $SU(3)$ case }
\label{sub2}
Next, we apply the model to the $SU(3)$ gauge group. In this case, there are two nontrivial
center elements in addition to the trivial center element
\begin{equation}
\mathbb Z(3)=\{z_0=1,z_1=e^{\frac{2\pi i}{3}},z_2=e^{\frac{4\pi i}{3}}\}.
\end{equation}
Since $z_1 = (z_2)^*$,  the vortex flux corresponding to $z_1$ is equivalent to an oppositely oriented vortex flux corresponding to $z_2$.
Therefore from Eq. (\ref {potential}), the static potential induced by all domains in $SU(3)$ gauge group is as the following:
\begin{equation}
V_r(R) =- \sum^{{ {L_d}/2 + R}}_{{x=-{L_d}/2}} \ln\Bigl\{(1-f_0-f_1-f_2) + f_0{\mathrm {Re}}\mathcal{G}_r[\alpha^{0}_C(x)] + f_1{\mathrm {Re}}\mathcal{G}_r[\alpha^{1}_C(x)]
+ f_2{\mathrm {Re}}\mathcal{G}_r[\alpha^{2}_C(x)]\Bigr\}, 
\end{equation}
where $f_1$, $f_2$, and $f_0$ are the probabilities that any given unit area is pierced by $z_1$ center vortex, $z_2$ center vortex, and the vacuum domain, respectively. As a result of 
Eq. (\ref {equal}), $f_1=f_2$ and ${\mathrm {Re}}\mathcal{G}_r[\alpha^{1}_R(x)]={\mathrm {Re}}\mathcal{G}_r[\alpha^{2}_R(x)]$. The free parameters are chosen as the same as  
subsection \ref{sub1}. The square ansatz for angles of center vortex and vacuum domain are:
\begin{equation}
         \Bigl((\alpha^1_C(x)\Bigr)^{2} = \frac{A_d}{ 2\mu} \left[
\frac{A}{ A_d} - \frac{A^2}{ A_d^2} \right]
                         + \left(\frac{4\pi}{\sqrt{3}}\frac{A}{ A_d}\right)^2,~~~
 \\
         \Bigl((\alpha^0_C(x)\Bigr)^{2} = \frac{A_d}{ 2\mu} \left[
\frac{A}{ A_d} - \frac{A^2}{ A^{2}_d} \right].                          
\end{equation}

\begin{figure}
\begin{center}
\resizebox{0.6\textwidth}{!}{
\includegraphics{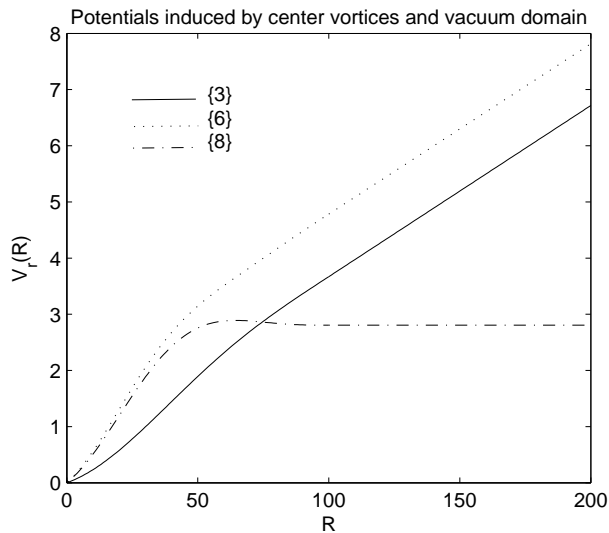}}
\caption{\label{T2}
The potentials between static sources using all domains for the fundamental, $\{6\}$, and adjoint representations of $SU(3)$ gauge group for the range $R\in [0,200]$. The 
free parameters are $L_{d}=100$, $f_{0}=0.03$, $f_{1}=0.01$ and $L^{2}_{d}/(2\mu)=4$. The string tensions agree qualitatively with Casimir scaling at intermediate distances, $R < 20$, 
and with  $N$-ality dependence at large distances.
}
\end{center}
\end{figure}

\begin{figure}
\begin{center}
\resizebox{0.6\textwidth}{!}{
\includegraphics{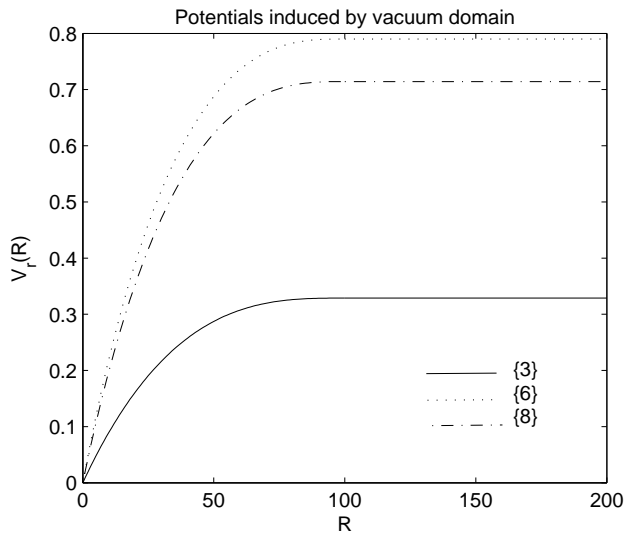}}
\caption{\label{T3}
The static potentials using the vacuum domain. A linear potential is observed for all representations.
}
\end{center}
\end{figure}

The static potentials $V_{r}(R)$, corresponding to all domains and vacuum domain, for the $\{3\}$ (fundamental), $\{6\}$ and $\{8\}$ (adjoint) representations for the range $R\in [0,200]$ 
are plotted in Figs. \ref{T2} and \ref{T3}. At intermediate distances, the potentials induced by all domains are linear in the range $R\in [0,20]$. It is clear from the plots that 
using only the vacuum domain, the potentials are also linear in the same interval. This phenomenon has already been observed for the SU($2$) gauge group as shown in Figs.
\ref{FIG2} and \ref{FIG3}.

\begin{figure*}[t]
\epsfxsize=8cm 
\epsffile{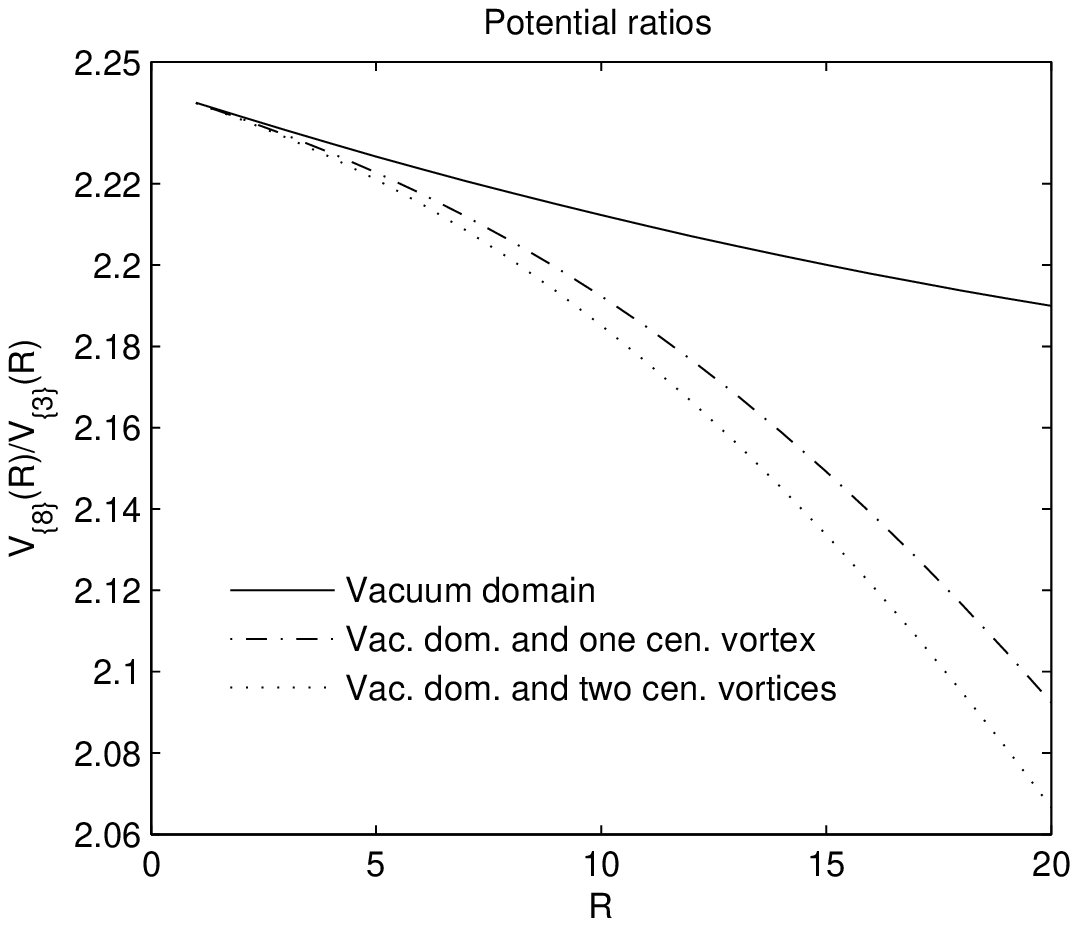} \hspace{0.0cm}
\epsfxsize=8cm 
\epsffile{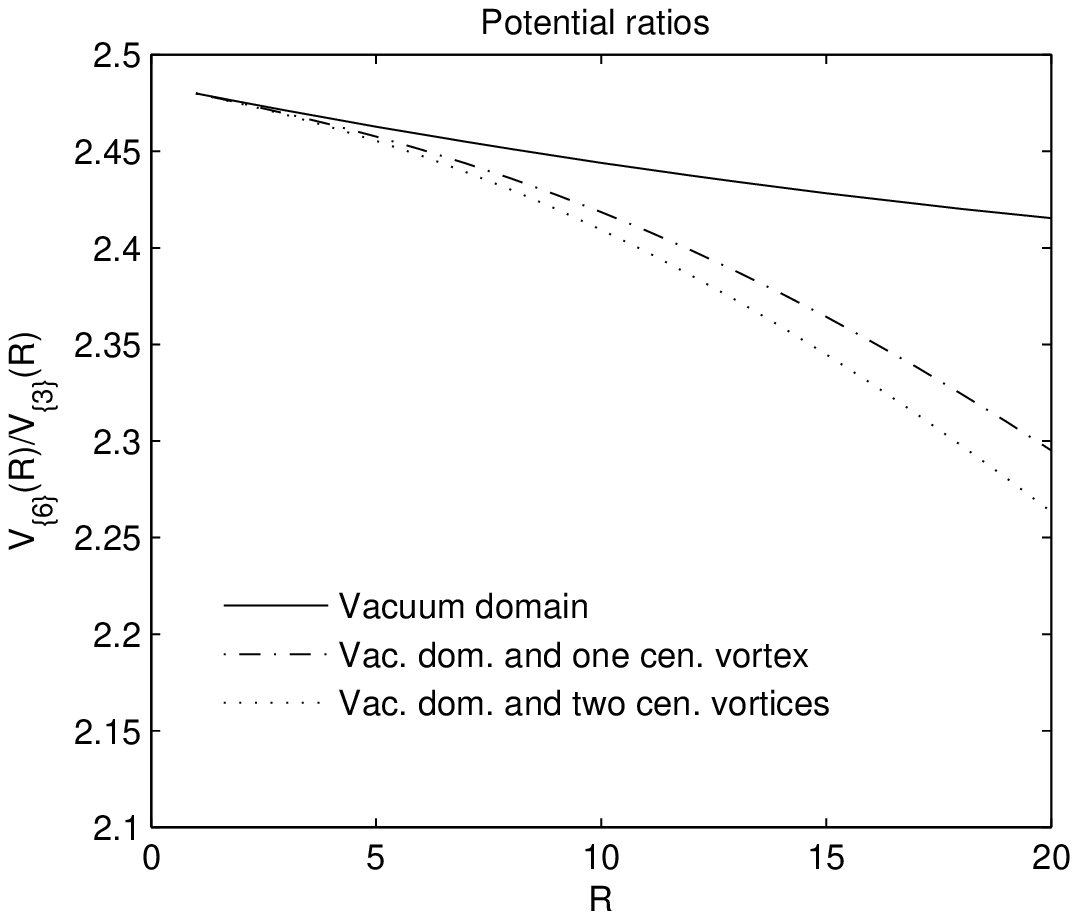} 
\caption{Left panel: ratios of ${V_{\{8\}}(R)}/{V_{\{3\}}(R)}$ are shown. Upper curve shows the contribution of the vacuum domain which violates very slowly from Casimir 
ratio compared with the contributions of the vacuum domain plus one center vortex (middle curve). The lower curve indicates the ratios when we use the vacuum domain plus two center 
vortices. By adding 
center vortices the slope of the ratios increases. Right panel: ratios of ${V_{\{6\}}(R)}/{V_{\{3\}}(R)}$ are shown. The results are the same as the adjoint case. 
The  free parameters are $L_{d}=100$, $f_{0}=0.03$, $f_{1}=0.01$ and $L^{2}_{d}/(2\mu)=4$.
}
\label{Y2}
\end{figure*}

\begin{figure}
\begin{center}
\resizebox{0.6\textwidth}{!}{
\includegraphics{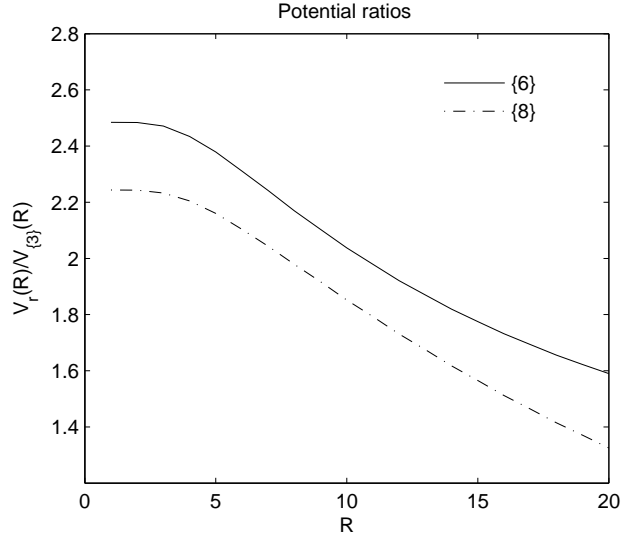}}
\caption{\label{B1}
Ratios of  $V_r(R)/V_{\{3\}}(R)$ induced by center vortex for adjoint and $\{6\}$ dimensional representations using old ansatz  for angle $\alpha_{C}(x)$.   
The  free parameters are $f_1=0.1$,~$a=0.05$, and~$b=4$. The potential ratios start out at the Casimir ratios but they reduce faster compared with 
the ratios obtained from the square ansatz. 
}
\end{center}
\end{figure}

Figure \ref{Y2} plots the potential ratios $V_{\{8\}}(R)/V_{\{3\}}(R)$ (left panel) and $V_{\{6\}}(R)/V_{\{3\}}(R)$ (right panel) for different contributions of domains.
 These potential ratios start out at the Casimir ratios: 
\begin{equation}
\frac{C_{\{8\}}}{C_{\{3\}}}=2.25, ~~~~~~~~~~~~~~~~~~~~~~~\frac{C_{\{6\}}}{C_{\{3\}}}=2.5.
\end{equation}
For the range $R \in [0,20]$, the potential ratios $V_{\{8\}}(R)/V_{\{3\}}(R)$ and  $V_{\{6\}}(R)/V_{\{3\}}(R)$ induced by the vacuum domain drop very slowly from $2.25$ and $2.5$ to about $2.19$ and $2.42$, respectively. Adding 
the contribution of the potential of the center vortex to the potential ratios $V_{\{8\}}(R)/V_{\{3\}}(R)$ and  $V_{\{6\}}(R)/V_{\{3\}}(R)$ obtained from the vacuum domain, the slope of the curve increases (the potential ratios change from $2.25$ and 
$2.5$ to about $2.1$ and $2.3$) and adding the contribution of the potential of the next center vortex, the slope of the curve increases again (the potential ratios change from $2.25$ and $2.5$ to 
about $2.07$ and $2.27$). 
On the other hand, Fig. \ref{B1} shows potential ratios using the old ansatz given in Eq. (\ref {Oansax}), with the  parameters $f=0.1$,~$a=0.05$, and~$b=4.$. The potential ratios 
 $V_{\{8\}}(R)/V_{\{3\}}(R)$ and  $V_{\{6\}}(R)/V_{\{3\}}(R)$ induced by center vortices drop from $2.25$ and $2.5$ to about $1.2$ and $1.6$ for the range $R \in [1,20]$, respectively. Therefore, the potential ratios 
drop slower using square ansatz compared with the old ansatz. 
From Fig. \ref{T2}, it is clear that at large distances, $R\geq 100$, the static potentials induced by all domains agree with $N$-ality. As shown in Fig. \ref{T3}, the potentials are screened at large distances where vacuum domain locates 
completely inside the Wilson loop. Therefore, to get the correct potentials at large distances, one has to use the center vortices and it is clear that the vacuum domains do not give the correct behavior. At large 
distances, a pair of gluon-anti gluon are popped out of the vacuum and combined with initial sources, and transform them into the lowest order representations in their class. Some 
examples are:
\begin{equation}
\{6\} \otimes \{8\}=\{\bar{3}\} \oplus \{6\} \oplus \{15\} \oplus \{24\},
\label{adjoint8}
\end{equation} 
\begin{equation}
\{8\} \otimes \{8\}= \{1\} \oplus \{8\} \oplus \{\bar{10}\} \oplus \{10\} \oplus \{27\}.
\label{8}
\end{equation}
In these examples  static sources in representations $\{6\}$ and $\{8\}$  are transformed into the lowest order
representation $\{\bar{3}\}$ and color singlet, respectively. Therefore, the slope of $\{6\}$ dimensional representation must be the same as the one for the fundamental representation and representation $\{8\}$ 
 must be screened. 
 
 In summary, using the vacuum domain only, the intermediate potentials agree better with Casimir scaling compared with the case when center vortices are using. In addition, square ansatz for the group factor is a better choice if one wants to see the Casimir scaling.

From this section, we conclude that for $SU(3)$, as well as $SU(2)$ gauge group, the potential ratios induced by vacuum domain agree better with Casimir scaling compared
with  the potential ratios induced by center vortices. In the next section, we argue about the reasons of these observations by studying the 
behavior of the potentials induced by vacuum domains and center vortices and the properties of the group factor $\mathcal{G}_r(\alpha^{(n)})$ 
for each case. 

\section{Interaction between the Wilson loop and center domains }
\label{5}
According to the center vortex theory, condensation of  non trivial center elements of the vacuum leads to confinement of static sources. Creation of a center vortex linked to a fundamental representation Wilson loop has the effect of multiplying the 
Wilson loop by an element of the gauge group center, $\it{i.e.}$ 
\begin{equation}
W_F(C)=Tr \big[U...U\big]\longrightarrow Tr \big[U...z...U\big].
\label{W2}
\end{equation}
One may claim that vacuum domains appear in the model as the result of interaction and combination of center vortices on the minimal area of the Wilson loop. In Sect. \ref{4}, we have 
shown that at intermediate distances, the static potential induced by only vacuum domains in $SU(N)$ gauge group, is linear and agrees better with Casimir scaling  
compared with  the static potential induced by center vortices. 
In the following subsection, we study the potentials between static sources and the behavior of the group factor $\mathcal{G}_r(\alpha^{(n)})$  
especially in $SU(2)$ gauge group to investigate the contribution of the center domains.

\subsection{Center domains in $SU(2)$}

\begin{figure}
\begin{center}
\resizebox{0.6\textwidth}{!}{
\includegraphics{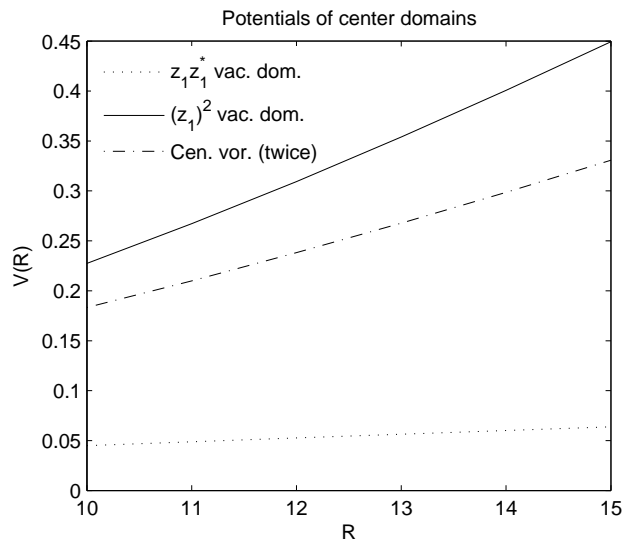}}
\caption{\label{POT}
The intermediate static potential induced by $(z_1)^2$ and $z_1z^*_1$ vacuum domains and the one which is obtained from twice of the static potential using center 
vortices. The extra positive potential 
energy of static potential induced by $(z_1)^2$ compared with the twice of the static potential obtained from center vortices shows that vortices with the fluxes in 
the same direction repel each other.  
On the other hand, extra negative energy of static potential induced by $z_1z^*_1$ compared with the twice of the static potential using center 
vortices shows that these vortices with the fluxes
in the opposite direction attract each other. The free parameters are chosen to be $L_{d}=100$, $f_{0}=0.01$, $f_{1}=0.01$ and $L^{2}_{d}/(2\mu)=4$. 
}
\end{center}
\end{figure}

\begin{figure}
\begin{center}
\resizebox{0.6\textwidth}{!}{
\includegraphics{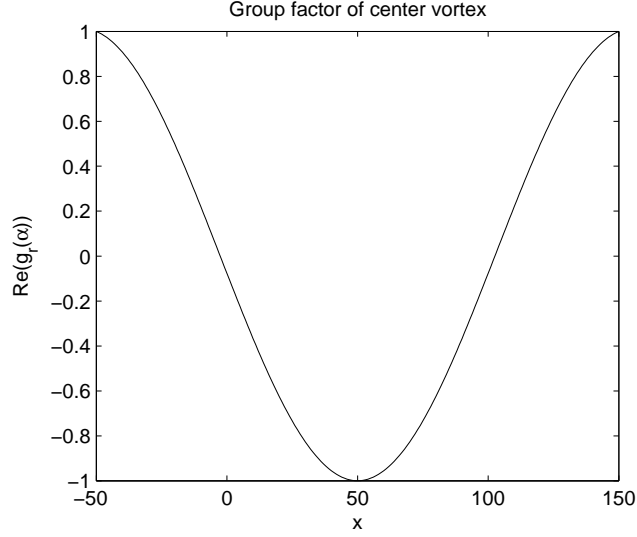}}
\caption{\label{2pi}
${\mathrm {Re}}(\mathcal{G}_{r})$ versus $x$ is plotted in the two dimensional representation ($j=1/2$) of the $SU(2)$ gauge group for a Wilson loop of $R = 100$. The free parameters are $L_{d}=100$ and $L^{2}_{d}/(2\mu)=4$. With the given parameters, the center vortex is completely located in the center of the Wilson loop at $x = 50$. Therefore the group factor changes 
between $1$ and $-1$.   
}
\end{center}
\end{figure}

For $SU(2)$ gauge group with the center group $\mathbb Z(2)=\{z_0=1,z_1=e^{\pi i}\}$, simultaneous creation of two similarly oriented center 
vortices linked to a Wilson loop leads to creation of vacuum domain i.e.
\begin{equation}
W_F(C)=Tr \big[U...U\big]\longrightarrow Tr \big[U...(z_1)^2...U\big],
\label{W3}
\end{equation}
and also simultaneous creation of two oppositely oriented center vortices linked to a Wilson loop, produces a vacuum domain as the following:
\begin{equation}
W_F(C)=Tr \big[U...U\big]\longrightarrow Tr \big[U...z_1z^*_1...U\big].
\label{W4}
\end{equation}
We recall that combining the center vortices fluxes has been studied in Ref. \cite{Engelhardt2000}, as well. For $SU(2)$ gauge group $(z_1)^2=z_1z^*_1=1$. Therefore, if the loop is large enough to contain two vortices, the vacuum domain is obtained. To understand the interaction between vortices, we study the potentials induced by vacuum domains and center vortices using the square ansatz. Figure \ref{POT} 
shows the static potentials of the fundamental representation, induced by vacuum domains corresponding to $(z_1)^2$ and $z_1z^*_1$ and 
center vortices. The potential energy induced by vacuum 
domains corresponding to two similarly oriented center vortices is larger than the twice of the potential energy induced by the center vortices. 
The extra positive energy may be interpreted as the interaction energy between center vortices constructing the vacuum domain. Therefore two 
vortices with the same flux orientations repel each other.   
On the other hand the potential energy induced by the vacuum domain corresponding to two oppositely oriented center vortices is less than 
the twice of the potential energy induced by the center vortices. Therefore an attraction occurs between two vortices with different flux 
orientations if they make a vacuum domain.   
Studying the group factors of the vacuum domains and the center vortices is also interesting \cite{Rafibakhsh2014}. 
 The group factor for the fundamental representation
of $SU(2)$ is obtained from Eq. (\ref {group factor}) and the Cartan of the $SU(2)$ gauge group:
 \begin{equation}
\mathcal{G}_{j=1/2}=cos(\frac{\alpha^{(n)}}{2}). 
 \end{equation}
For the fundamental representation of $SU(2)$ gauge group, when the center vortex is
completely contained within the Wilson loop,
\begin{equation}
\exp(i\vec{\alpha}^{(1)}\cdot\vec{\mathcal{H}}_{Fun})= z_1I.
\end{equation} 
Using the Cartan generator of $SU(2)$, the maximum value of the angle $\alpha^{(1)}_{max}$ for the fundamental representation is equal to 
$2\pi$. Figure \ref{2pi} plots $\mathcal{G}_r(\alpha^{(n)})$ versus $x$ for a Wilson loop with $R=100$ for the fundamental representation of $SU(2)$ using square ansatz. The Wilson 
loop legs are located at $x = 0$ and $x = 100$. When the center vortex overlaps the minimal area of the Wilson loop, it affects the loop. The group 
factor interpolates smoothly from $-1$, when the vortex core is located entirely within the Wilson loop, to $1$, when the core is entirely outside the loop. The interaction between center vortices is not considered.  

\begin{figure*}[t]
\epsfxsize=8cm 
\epsffile{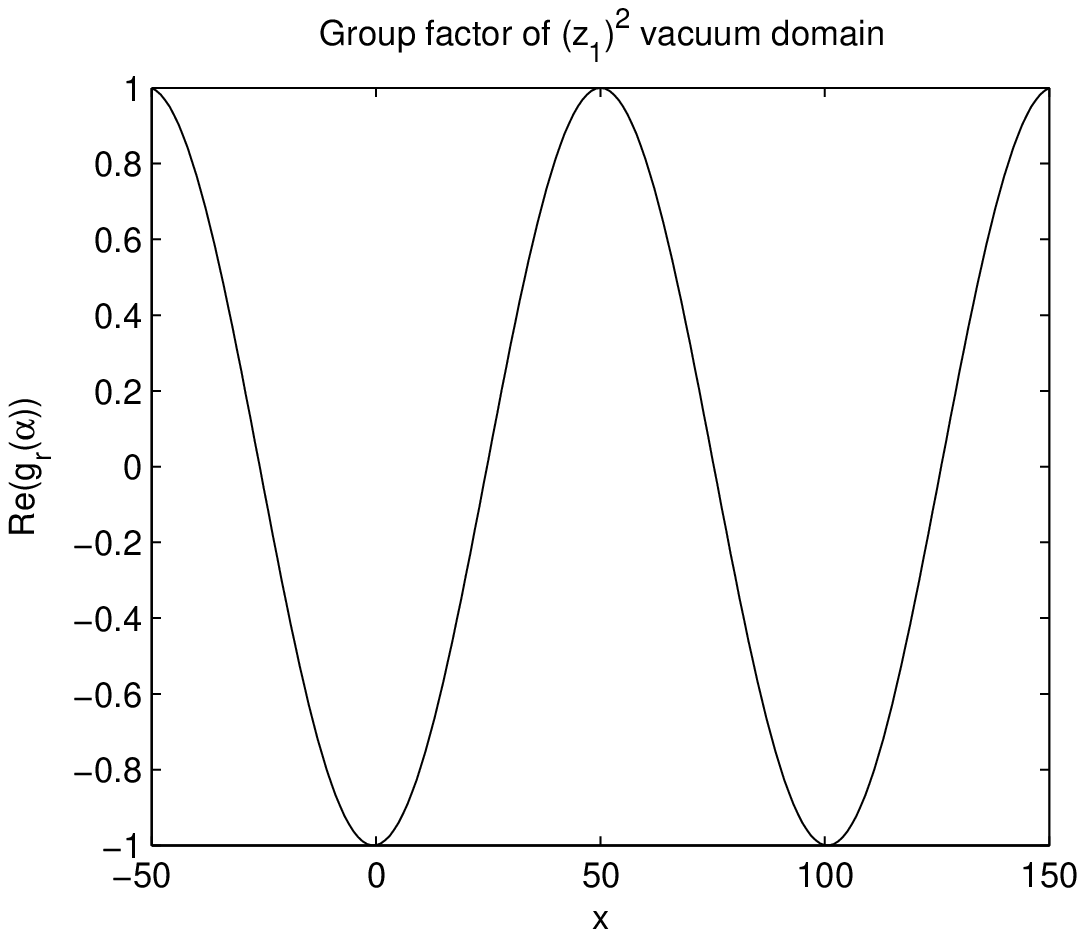} \hspace{0.0cm}
\epsfxsize=8cm 
\epsffile{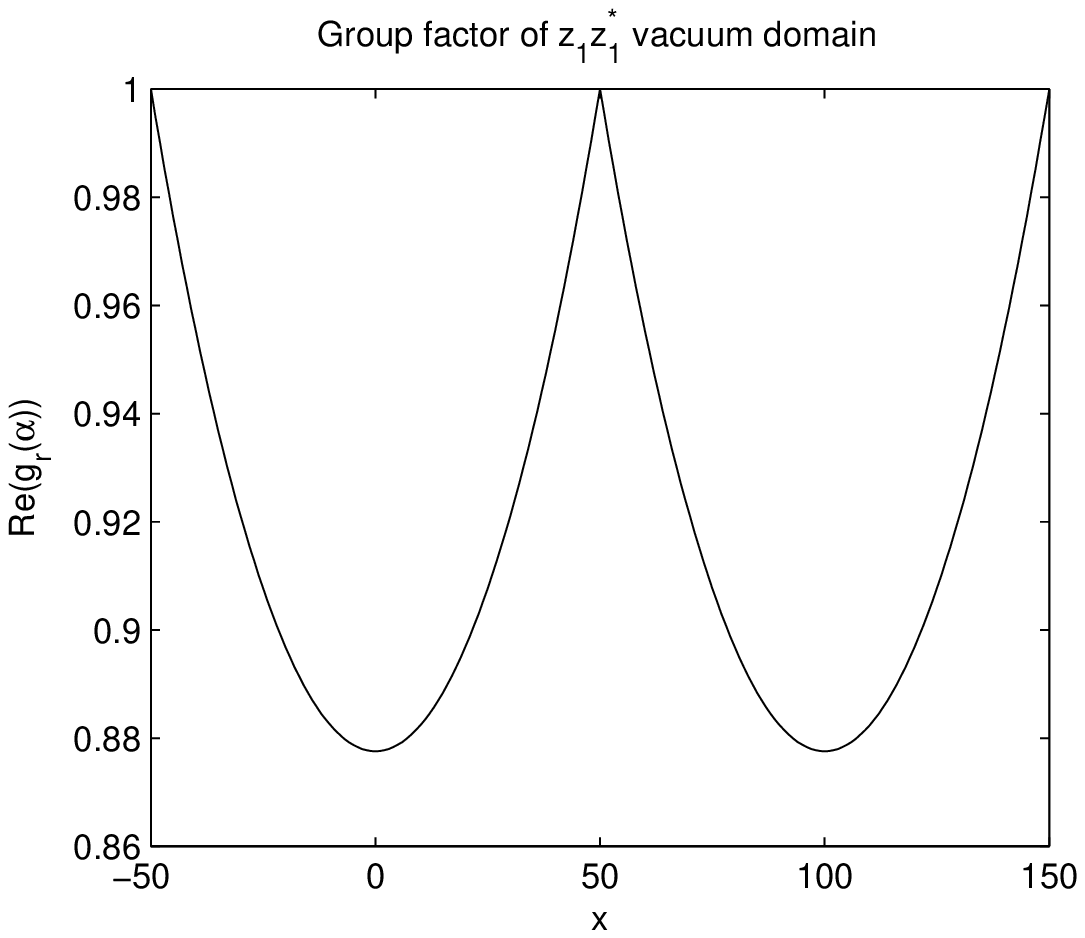} 
\caption{${\mathrm {Re}}(\mathcal{G}_{r})$ induced by a vacuum domain versus $x$ is plotted for the two dimensional representation ($j=1/2$) of 
the $SU(2)$ gauge group for $R = 100$. The group factor of $(z_1)^2$ vacuum domain changes between $1$, where the core of the vacuum 
domain locates completely in the Wilson loop (at $x = 50$), and $-1$ where half flux of vacuum domain locates in the Wilson loop (at $x = 0$ or $x = 100$). When $50\%$ 
of the vacuum domain core locates in the Wilson loop, the flux inside the loop is equivalent to the center vortex flux. Since two similarly oriented
 vortices repel each other, the magnetic flux in each vortex conserves and we observe $-1$ (corresponding to the $SU(2)$ center vortex) shown in the plot of the group factor of 
vacuum domain (left panel). The group factor of $z_1z^*_1$ vacuum domain changes between $1$, where the core of the vacuum 
domain locates completely in the Wilson loop (at $x = 50$), and about $0.9$ where half flux of the vacuum domain locates in the loop (at $x = 0$ or $x = 100
 $). One can argue that since two oppositely center vortices attract each other, the cores of two oppositely oriented vortices overlap 
 each other and some part of magnetic flux in each vortex is annihilated. Therefore ${\mathrm {Re}}(\mathcal{G}_{r})$ at $x = 0$ or $x = 100$ is close to one (about $0.9$) (right panel). 
 The free parameters are $L_{d}=100$ and $L^{2}_{d}/(2\mu)=4$. 
}
\label{Y6}
\end{figure*}

\begin{figure}
\begin{center}
\resizebox{0.3\textwidth}{!}{
\includegraphics{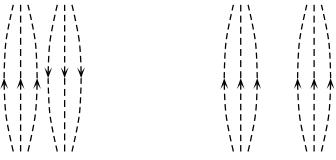}}
\caption{\label{Q1}
The figure schematically shows vacuum domains. Center vortices of the $(z_1)^2$ vacuum domain (right panel) repel each other while the 
ones in constructing $z_1z^*_1$ vacuum domain (left panel) attract each other and the cores of center vortices overlap each other. 
}
\end{center}
\end{figure}

 On the other hand when the vacuum domain corresponding to $(z_1)^2$ is located completely inside the Wilson loop, 
 \begin{equation}
\exp(i\vec{\alpha}^{(0)}\cdot\vec{\mathcal{H}}_{Fun})=(z_1)^2I.
\end{equation}  
 Therefore, the maximum value of the flux profile $\alpha^{(0)}_{max}$ for the fundamental representation is equal to $4\pi$.
 Figure \ref{Y6} (left) plots $\mathcal{G}_r(\alpha^{(n)})$ versus $x$ for $R = 100$ for the fundamental representation of $SU(2)$. The cross section of the vacuum domain is a $L_d \times L_d$ square and $L_d=100$. If the center of the vacuum domain is placed
at $x = 0$ or $x = 100$, $50\%$ of the maximum flux enters the Wilson loop 
 
 \begin{equation}
 \mathrm {Re}(\mathcal{G}_{r})= \mathrm {Re}\frac{1}{d_r}\mathrm{Tr}\left(\exp\left[\frac{i}{2}\alpha^{max}\mathcal{H}_3^{f}\right]\right)=-1,
\end{equation} 
 where this is equal to the minimum of the group factor of $SU(2)$ center vortices. In other words, when $50\%$ of the flux of the vacuum domain locates within the loop, the flux of center vortex is obtained. Since two vortices in the $(z_1)^2$ vacuum domain repel each other, the magnetic flux in each vortex conserves and we observe $-1$ for the group factor of vacuum domain when half flux of the vacuum domain locates in the Wilson loop. Figure \ref{Q1} (right) schematically shows the $(z_1)^2$ vacuum domain. 
 Next, we study the $z_1z^*_1$ vacuum domain when it is located completely inside the Wilson loop,
\begin{equation}
\exp(i\vec{\alpha}^{(0)}\cdot\vec{\mathcal{H}}_{Fun})=z_1z^*_1I.
\end{equation}  
\begin{figure*}[t]
\epsfxsize=8cm 
\epsffile{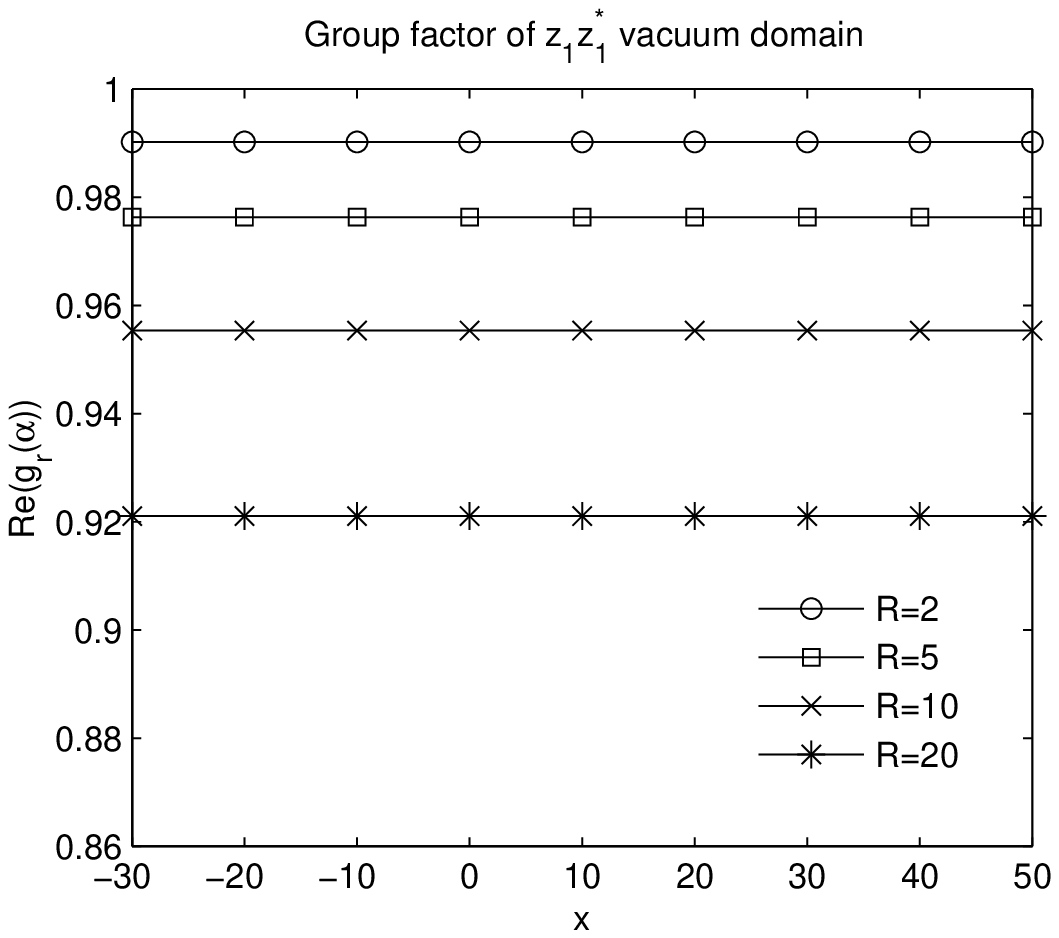} \hspace{0.0cm}
\epsfxsize=8cm 
\epsffile{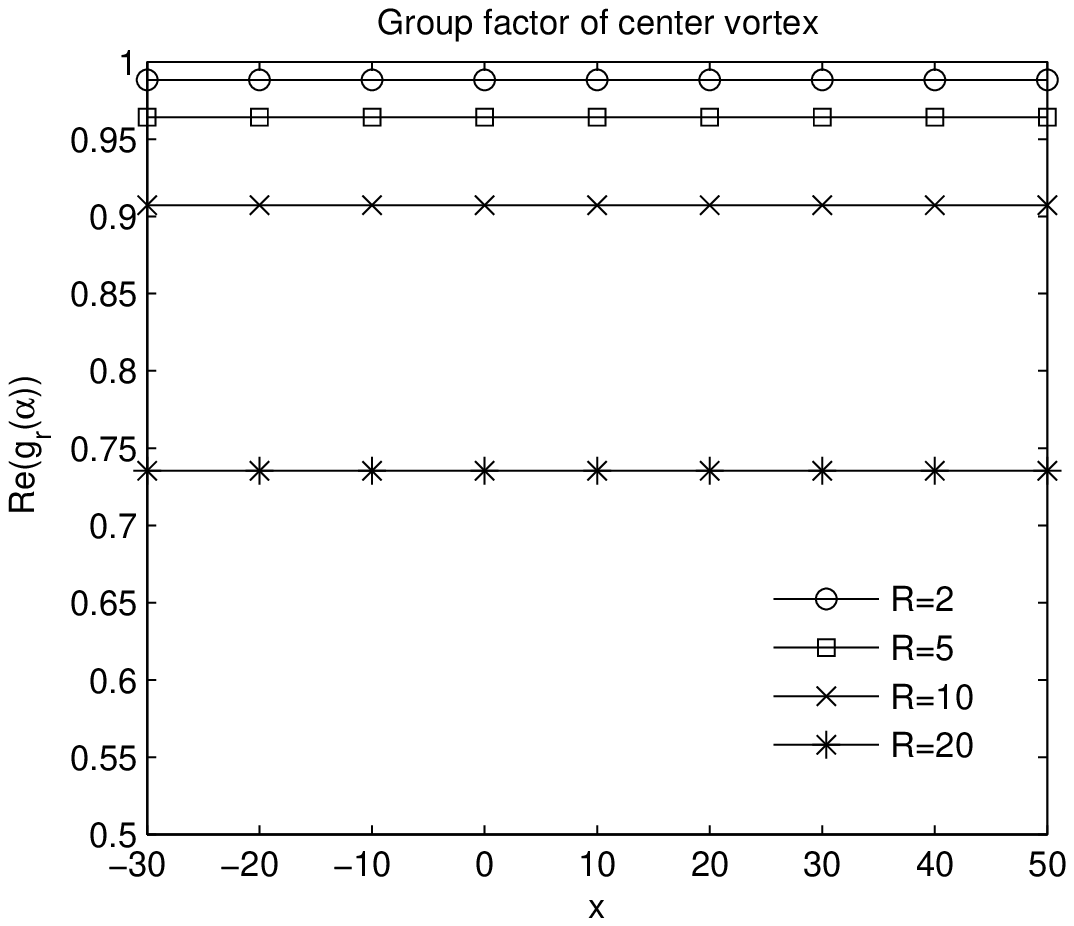} 
\begin{center}
\epsfxsize=8cm 
\epsffile{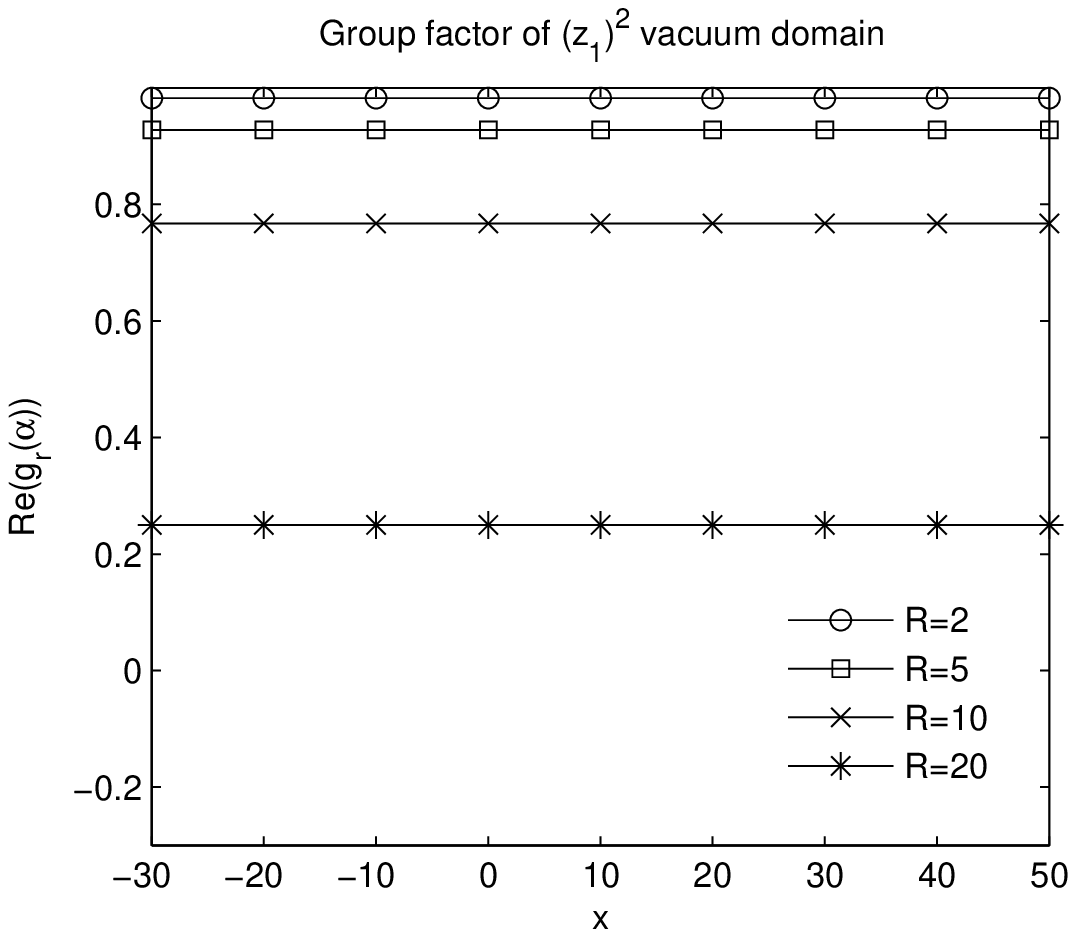}
\end{center}
\caption{${\mathrm {Re}}(\mathcal{G}_{r})$ obtained from center domains versus $x$ for different sizes of Wilson loops (different $R$) in the Casimir scaling regime. 
It is plotted for the two dimensional representation ($j=1/2$) of the $SU(2)$ gauge group. In this regime, the group factor of
center vortices changes slowly from $1$ to $0.75$. For the same regime, the group factor of $z_1z^*_1$ vacuum domain  changes from $1$ to $0.92$ which is very slower compared with the one obtained from center vortices. Also the group factor obtained from $(z_1)^2$ vacuum domains changes very fast from $1$ to $0.2$ compared with the one obtained from center vortices. Therefore as discussed in the text, by adding the contribution of the $z_1z^*_1$ vacuum domain in the potential obtained from center vortices, the length of Casimir scaling regime increases and by adding the contribution of the $(z_1)^2$ vacuum domain in the potential obtained from center vortices, the length of Casimir scaling regime decreases.   
}
\label{Z5}
\end{figure*}
Therefore, the maximum value of the flux profile $\alpha^{(0)}_{max}$ for the fundamental representation is zero and in this case $\mathrm {Re}(\mathcal{G}_{r})=1$.
 Fig. \ref{Y6} (right) plots ${\mathrm {Re}}(\mathcal{G}_{r})$ versus $x$ for $R = 100$ for the $z_1z^*_1$ vacuum domain in the fundamental representation of $SU(2)$. If the 
center of the vortex core is placed
at $x = 0$ or $x = 100$, $50\%$ of the maximum flux enters the Wilson loop and the value of the group factor is about $0.9$. Since two oppositely 
oriented vortices of the $z_1z^*_1$ vacuum domain attract each other, the cores of two oppositely oriented 
vortices overlap each other and some part of the magnetic flux in each vortex is annihilated. Figure \ref{Q1} (left) schematically represents  
the $z_1z^*_1$ vacuum domain. As a result, the magnetic flux of the center vortices does not conserve and we do not observe -$1$ (corresponding to 
the $SU(2)$ center vortex) for the group factor of the vacuum domain when half flux of the vacuum domain locates inside the Wilson loop. 

 Now, we discuss the effect of adding the contributions of the vacuum domains corresponding to $(z_1)^2$ and $z_1z^*_1$ to the potential induced by center vortices. According to Figs. \ref{Y1} and \ref{Y2}, the potential ratios start out at the ratios of the corresponding Casimirs. Therefore for small size loops ($R \approx 1,2$) where $\alpha_{c}$ is also small ($\alpha_{c} \approx 0$), the potentials strongly agree with the Casimir scaling. As a result, for small size loops, the group factor is close to one $\it{i.e.}$ ${\mathrm {Re}}(\mathcal{G}_{r}) \approx 1$. So if the group factors in medium size loops ($R<20$) change very slowly, the potential ratios drop smoothly from Casimir ratios. 
 
 A comparison between group factors obtained from different domains for the two dimensional representation ($j=1/2$) of the $SU(2)$ gauge group is done by plotting Fig. \ref{Z5} for the Casimir scaling regime. The value of the group factor obtained from center vortices changes smoothly from $1$ to $0.75$ for $R<20$. In the same range of distances, the value of the group factor obtained from $z_1z^*_1$ vacuum domains changes from $1$ to $0.92$ where the changing rate is slower than the one obtained from center vortices. Therefore the magnetic flux of $z_1z^*_1$ vacuum domain linked to different medium size loops ($R<20$) is approximately close to zero and their group factors is close to one. Since the group factor obtained from $z_1z^*_1$ vacuum domains changes slower than the one obtained from center vortices, therefore the potential ratios obtained from $z_1z^*_1$ vacuum domains violate from Casimir ratios slower than the one obtained from the center vortices. Also the value
  of group factor obtained from $(z_1)^2$ vacuum domains changes very fast from $1$ to $0.2$. Since the group factor obtained from $(z_1)^2$ vacuum domains changes faster than the one obtained from center vortices, therefore the potential ratios obtained from $(z_1)^2$ vacuum domains violate quickly from Casimir ratios.
 
 In summary, in the intermediate regime, the potential ratios obtained from $z_1z^*_1$ vacuum domain drop slower than the one induced by center vortices, and the potential ratios obtained from $(z_1)^2$ vacuum domain drop faster than the one obtained from center vortices. Therefore by adding the contribution of the $z_1z^*_1$ vacuum domain to the potential obtained from center vortices, the length of the Casimir scaling regime increases and by adding the contribution of the $(z_1)^2$ vacuum domain to the potential induced by center vortices, the length of Casimir scaling regime decreases. The above discussion can explain why the length of Casimir scaling is increased in Fig. \ref{FIG2}. It is obvious that both vacuum structures $z_1z^*_1$ and $(z_1)^2$ have contribution in the potential but it seems that $z_1z^*_1$ has a dominant role in increasing the Casimir regime. One can use the same arguments for $SU(3)$ gauge group for explaining the potentials induced by the domains in Fig. \ref{T2}.         

\subsection{Comparison between $SU(N)$ and $G(2)$ gauge groups}
\begin{figure}
\begin{center}
\resizebox{0.6\textwidth}{!}{
\includegraphics{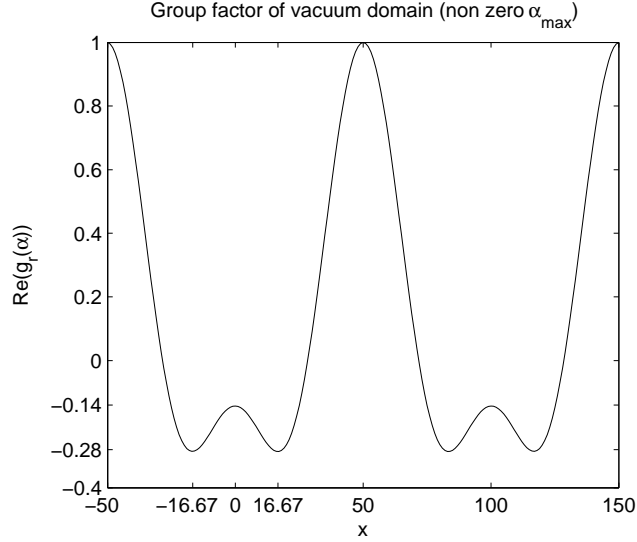}}
\caption{\label{F1}
${\mathrm {Re}}(\mathcal{G}_{r})$ induced by vacuum domain versus $x$ is plotted in the $7$ dimensional (fundamental) representation of the 
$G(2)$ gauge group for $R = 100$. The group factor has local extremums $-0.14$ and $-0.28$ corresponding to $SU(2)$ and $SU(3)$ subgroups 
of $G(2)$, respectively. In the some nearby plaquettes of the legs of timelike, the fluxes inside the loop is equivalent to center vortex fluxes of 
$SU(2)$ and $SU(3)$ subgroups of $G(2)$. One can argue that when partial flux of vacuum domain locates into the Wilson loop as if fluxes of 
center vortices of subgroups locate into the loop. The extremums of vacuum domain group factor in $SU(N)$ gauge group 
correspond to non trivial center elements of groups (agents of confinement) while the ones in $G(2)$ gauge group correspond to center elements of 
subgroups. The selected free parameters are $L_{d}=100$ and $L^{2}_{d}/(2\mu)=4$. 
}
\end{center}
\end{figure}

As argued, for $SU(N)$ gauge groups which have non trivial center elements, the group factor of vacuum domain changes between $1$ and non 
trivial center elements of the gauge group. It is interesting to compare the 
behavior of the group factors of $SU(N)$ gauge group and $G(2)$ gauge group which has only trivial center element $z_0=1$. For $G(2)$ gauge group, 
a linear regime in agreement with Casimir scaling is observed from both lattice gauge theory \cite{Greensite2007,Olejnik2008} and  
domain model \cite{Deldar2012}. The entire $G(2)$ group can be covered by six $SU(2)$ subgroups \cite{Cossu2007dk}. Three of them, the non reducible ones, generate an $SU(3)$ subgroup of $G(2)$ which is seven dimensional and reducible. The representations of the remaining 
three $SU(2)$ subgroups are seven dimensional, but they are reducible. The center elements of the $SU(3)$ and $SU(2)$ subgroups of $G(2)$ in the 
fundamental representation are given by 
\begin{equation}
\label{Z}
Z_a^{SU(3)}= \left(\begin{array}{ccc} z_aI_{3\times3} & 0 & 0 \\0 &1 &0 \\ 0 & 0 & z_a^*I_{3\times3}
\end{array} \right),~~~~~~Z_a^{SU(2)}= \left(\begin{array}{ccc} z_aI_{2\times2} & 0 & 0 \\0 &z_aI_{2\times2} &0 \\ 0 & 0 & I_{3\times3}
\end{array} \right),
\end{equation}
where $I$ is the unit matrix, and $z_a\in\{z_0=1,z_1=e^{\frac{2\pi i}{3}},z_2=e^{\frac{4\pi i}{3}}\}$ for the $SU(3)$ subgroup center elements, and 
$z_a\in\{z_0=1,z_1=e^{\pi i}\}$ for the $SU(2)$ subgroup center elements.
We discussed the possible reasons of observing the confined potential at intermediate distances in our previous article \cite{Deldar2014}. We studied
 ${\mathrm {Re}}(\mathcal{G}_{r})$ for the $G(2)$ gauge group. Using ansatz given in Eq. (\ref {Sansax}), Fig. \ref{F1} plots ${\mathrm {Re}}(\mathcal{G}_r(\alpha^{(n)}))$ versus 
$x$ for the $7$ dimensional (fundamental) representation of $G(2)$ gauge group for $R=100$ . The timelike legs of the Wilson loop are located 
at $x = 0$ and $x = 100$. The group factor of the vacuum domain changes between $1$ and the non trivial center elements of the $SU(2)$ and $SU(3)$ 
subgroups as the following:    
\begin{equation}
\label{BB}
 {min}[{\mathrm {Re}}\mathcal{G}_{r}
(\alpha(x) )]_ {SU(2)}=
\frac{1}{7}{\mathrm {Re}}\mathrm{Tr}\left(\begin{array}{ccc} e^{i\pi}I_{2\times2} & 0 & 0 \\0 &e^{i\pi}I_{2\times2} &0 \\ 0 & 0 & I_{3\times3}
\end{array} \right)\\ 
=-0.14,
\end{equation}
\begin{equation}
\label{BBB}
 {min}[{\mathrm {Re}}\mathcal{G}_{r}
(\alpha(x))]_ {SU(3)}=\frac{1}{7}{\mathrm {Re}}\mathrm{Tr}\left(\begin{array}{ccc} e^\frac{i2\pi}{3}I_{3\times3} & 0 & 0 \\0 &1 &0 \\ 0 & 0 & e^{-\frac {i2\pi}{3}}I_{3\times3}
\end{array} \right)\\
=-0.28.
\end{equation}   

For $SU(N)$ gauge group, we have argued that vacuum domains may appear as a results of simultaneously creation of some center vortices linked to the Wilson loop. Therefore 
the foot print of center vortices has been observed in extremums of the vacuum domain group factor. Although $G(2)$ gauge 
group does not have any center vortex but the extremums of the vacuum domain group factor have been related to the subgroups of $G(2)$. One may argue that 
the $G(2)$ vacuum is filled with center vortices of the subgroups.  
Simultaneous creation of three similarly oriented center vortices of the $SU(3)$ subgroup linked to a Wilson loop may give a vacuum 
domain:
\begin{equation}
W_F(C)=Tr \big[U...U\big]\longrightarrow Tr \big[U...(Z_1^{SU(3)})^3...U\big].
\label{W5}
\end{equation}
Simultaneous creation of two similarly oriented center vortices of $SU(2)$ subgroup linked to a Wilson loop leads to creation of vacuum domain, as well: 
\begin{equation}
W_F(C)=Tr \big[U...U\big]\longrightarrow Tr \big[U...(Z_1^{SU(2)})^2...U\big].
\label{W6}
\end{equation}
Now we discuss about the local extremums of the group factor of the vacuum domain in $G(2)$ gauge group in Fig. \ref{F1}. Since a combination of three $Z_1^{SU(3)}$ linked to a Wilson loop leads to the vacuum domain, therefore we expect that one third of the vacuum domain gives the $Z_1^{SU(3)}$ flux. In the vicinity of the left timelike leg of the Wilson loop, there are two minimums at $x=-16.67$ and $x=16.67$. If the center of the vacuum domain core is placed at $x=-16.67$, because the size of the core of the domain is equal to $100$, about one third of the vacuum domain enters the Wilson loop and the group factor is obtained as the following:
 \begin{equation}
 \mathrm {Re}(\mathcal{G}_{r})= \mathrm {Re}\frac{1}{d_r}\mathrm{Tr}\left(\exp\left[\frac{i}{3}\alpha^{max}\mathcal{H}_8^{f}\right]\right)=-0.28.
\end{equation}
This value agree with Eq. (\ref {BBB}). Therefore at $x=-16.67$ where one third of the vacuum domain locates in the Wilson loop, the partial flux is equal to the flux of $Z_1^{SU(3)}$ vortex. 
If the center of the vacuum domain core is placed at $x=16.67$, about two third of the vacuum domain enters the Wilson loop and the group factor is obtained as the following:
 \begin{equation}
 \mathrm {Re}(\mathcal{G}_{r})= \mathrm {Re}\frac{1}{d_r}\mathrm{Tr}\left(\exp\left[\frac{2i}{3}\alpha^{max}\mathcal{H}_8^{f}\right]\right)=-0.28.
\end{equation} 
At $x=16.67$, where two third of the vacuum domain locates in the loop, one can expect that the partial flux located in the Wilson loop is different from the partial flux at  $x=-16.67$. But two third of the vacuum domain is equal to $(Z_1^{SU(3)})^2$ flux which is as the following:
 \begin{equation}
Z_1^{SU(3)}\times Z_1^{SU(3)}=\left(\begin{array}{ccc} e^\frac{i2\pi}{3}I_{3\times3} & 0 & 0 \\0 &1 &0 \\ 0 & 0 & e^{-\frac {i2\pi}{3}}I_{3\times3}\end{array} \right)\times \left(\begin{array}{ccc} e^\frac{i2\pi}{3}I_{3\times3} & 0 & 0 \\0 &1 &0 \\ 0 & 0 & e^{-\frac {i2\pi}{3}}I_{3\times3}\end{array} \right)=(Z_1^{SU(3)})^*.
\end{equation}
This means that the $(Z_1^{SU(3)})^2$ flux is equivalent to an oppositely oriented vortex flux of $Z_1^{SU(3)}$ $\it{i.e.}$  $(Z_1^{SU(3)})^*$ vortex. Therefore we observe flux of $Z_1^{SU(3)}$ at $x=16.67$, where two third of the vacuum domain locates in the loop. 

On the other hand since a combination of two $Z_1^{SU(2)}$ linked to a Wilson loop may create a vacuum domain, therefore we expect that half flux of the vacuum domain leads to $Z_1^{SU(2)}$ flux. If the center of the vacuum domain core is placed at $x=0$, about half of the vacuum domain enters the Wilson loop and the group factor is obtained:
 \begin{equation}
 \mathrm {Re}(\mathcal{G}_{r})= \mathrm {Re}\frac{1}{d_r}\mathrm{Tr}\left(\exp\left[\frac{i}{2}\alpha^{max}\mathcal{H}_8^{f}\right]\right)=-0.14.
\end{equation}
This value agree with Eq. (\ref {BB}). Therefore at $x=0$ where half of the vacuum domain locates in the Wilson loop, the partial flux is equal to the flux of $Z_1^{SU(2)}$ vortex. One can do the same discussion in the vicinity of the right timelike leg of the Wilson loop.

As a result, in comparison with $SU(N)$ Yang-Mills theory where local extremums correspond to non trivial center elements of the gauge group, the local extremums for $G(2)$ correspond to the non trivial center elements of the $SU(2)$ and $SU(3)$ subgroups. In other word, the vacuum domain in $SU(N)$ depends on the center vortices of the gauge groups and in $G(2)$ depends on the center vortices of its subgroups.

\section{Conclusions}
\label{6}
Applying thick vortex model, which contains the vacuum domains, to the $SU(2)$ and $SU(3)$ gauge groups and using square ansatz for angle 
$\alpha_{C}(x)$, we show that the static potentials of various representations grow linearly at 
intermediate distances and agree with $N$-ality at large distances. We compute Casimir ratios for $SU(2)$ and $SU(3)$ color sources at intermediate distances and we show that they are qualitatively in better agreement with Casimir ratios when using the square ansatz rather than the old ansatz for angle $\alpha_{C}(x)$. We also study the contributions
of the vacuum domain and center vortices to the static potentials. Our results for $SU(2)$ and $SU(3)$ gauge groups show that the potential ratios obtained from the vacuum domain agree better with Casimir scaling than the potential ratios obtained from center vortices. We discuss about the reason of these observations by studying the potentials and the  
 group factor $\mathcal{G}_r(\alpha^{(n)})$. The group factor plays an important role in the potential between quarks.  According to vortex theory, the vacuum of QCD is filled with non trivial center vortices. One can construct vacuum domain by simultaneously creation of 
the center vortices linking the Wilson loop. We have discussed about the $z_1z^*_1$ and $(z_1)^2$ vacuum domains in the $SU(2)$ gauge group. It seems that two oppositely center vortices in $z_1z^*_1$ vacuum domain attract each other and two similarly oriented
 vortices in $(z_1)^2$ vacuum domain repel each other. Therefore in $SU(2)$ gauge group, the group factor of $(z_1)^2$ vacuum domain changes 
between $1$ and non trivial center elements of the gauge group and the one of $z_1z^*_1$ vacuum domain changes 
between $1$ and $0.9$. Since the potential ratios start out at the ratios of the corresponding Casimirs, therefore for small size loops ($R \approx 1,2$) where the group factor is close to one, the potentials strongly agree with the Casimir scaling.  In the intermediate regime, the potential ratios obtained from $z_1z^*_1$ vacuum domain drop slower than the one obtained from center vortices and the potential ratios obtained from $(z_1)^2$ vacuum domain drop faster than the one obtained from center vortices. Therefore the length of Casimir scaling regime increases by adding the contribution of the $z_1z^*_1$ vacuum domain to the potential induced by center vortices. On the other hand, by adding the contribution of the $(z_1)^2$ vacuum domain to the potential, the length of Casimir scaling regime decreases.    

Comparison between the behavior of the group factor in $SU(N)$ gauge group with non trivial 
center elements, and $G(2)$ gauge group with no non trivial center element is done, as well. In $SU(N)$ gauge groups, the group factor 
changes between $1$ and center vortices of the group, but in $G(2)$ gauge group it changes between $1$ and center vortices of $SU(2)$ and $SU(3)$ 
subgroups. One can argue that the $SU(2)$ and $SU(3)$ subgroups have
dominant roles in confinement regime in $G(2)$ gauge group.  

\appendix
\section{ Cartan generators }
One can obtain $\mathcal{H}_a^r$, the Cartan generator in $r$ representation, by using the tensor method. If $\{X_r^i;i=1,2,...,d_r\}$ is defined as the basis vector for the representation $r$ with dimension $d_r$, the elements of $\mathcal{H}_a^r$ can be computed by:
\begin{equation}
\mathcal{H}_a^rX_r^i=\sum_{j=1}^{d_r} C_{ij} X_r^j.
\end{equation}
Using the explicit basis, $X_r^i$, for representation $r$, $C_{ij}$, which are the coefficients, can be computed. 
The generators of higher representation can be obtained by lower representations \cite{georgi}:
\begin{equation}
{\big(\mathcal{H}_a^{\{D_1\}\otimes \{D_2\}}\big)}_{ix,iy}=\big (\mathcal{H}_a^{\{D_1\}}\big) \delta_{xy}+\delta_{ij} \big (\mathcal{H}_a^{\{D_2\}}\big).
\label{higher}
\end{equation}
$\mathcal{H}_a^{D_i}$s are the group generators for representations $\{D_1\}$, $\{D_2\}$ and $\{D_1\}\otimes \{D_2\}$.
First, we calculate the Cartan generators for the $SU(3)$ representations $\{6\}$ and $\{8\} (adj)$. $v^i$ and $u^i$, ${i,j=1,...,3}$ are considered as the basis vectors for the quarks in fundamental representation. Therefore the basis tensor for $\{6\}$ representation which is given by $\{3\}\otimes \{3\}$, is as the following:
\begin{equation}
V^{ij}=\frac{1}{2}\big(v^i u^j+v^ju^i\big).
\label{base6}
\end{equation}
The six independent states are:
\begin{eqnarray}
\begin{array}{llll}
X^1_6&=V^{11}=v^1u^1,\\
X^2_6&=V^{12}=\frac{1}{2}(v^1u^2+v^2u^1),\\
X^3_6&=V^{13}=\frac{1}{2}(v^1u^3+v^3u^1),\\
X^4_6&=V^{22}=v^2u^2,\\
X^5_6&=V^{23}=\frac{1}{2}(v^2u^3+v^3u^2),\\
X^6_6&=V^{33}=v^3u^3.\\
\end{array} \
\end{eqnarray}
From Eq. (\ref {higher}) and the above basis tensor, the $\mathcal{H}_a^6$ generators ($a=3,8$) are calculated as the following:
\begin{equation}
\mathcal{H}_3^6=diag\big(1,0,\frac{1}{2},-1,-\frac{1}{2},0\big),
\end{equation}
\begin{equation}
\mathcal{H}_8^6=\frac{1}{\sqrt3}diag\big(1,1,-\frac{1}{2},1,-\frac{1}{2},-2\big).
\end{equation}
Also the basis tensor for $\{8\}$ representation, given by $\{3\} \otimes \{\bar3$\}, is:
\begin{equation}
U^i_j=v^iu_j-\frac{1}{3}\delta^i_jv^ku_k.
\label{base8}
\end{equation}
The eight independent states are:

\begin{eqnarray}
\begin{array}{llll}
X^1_8&=U^1_1=v^1u_1-\frac{1}{3}v^ku_k,\\
X^2_8&=U^1_2=v^1u_2,\\
X^3_8&=U^1_3=v^1u_3,\\
X^4_8&=U^2_1=v^2u_1,\\
X^5_8&=U^2_2=v^2u_2-\frac{1}{3}v^ku_k,\\
X^6_8&=U^2_3=v^2u_3,\\
X^7_8&=U^3_1=v^3u_1,\\
X^8_8&=U^3_2=v^3u_2.\\
\end{array} \
\end{eqnarray}
From Eq. (\ref {higher}) and the above basis tensor, the $\mathcal{H}_a^8$ generator are calculated as the following:
\begin{equation}
\mathcal{H}_3^8=diag\big(0,1,\frac{1}{2},-1,0,-\frac{1}{2},-\frac{1}{2},\frac{1}{2}\big),
\end{equation}
\begin{equation}
\mathcal{H}_8^8=\frac{3}{2\sqrt3}diag\big(0,0,1,1,0,0,1,-1,-1\big).
\end{equation}
Next, we obtain Cartan generator of $G(2)$ gauge group using $SU(3)$ Cartan generators. The $\{7\}$ (fundamental) dimensional representation of $G(2)$ gauge group under $SU(3)$ subgroup transformations decomposes into
\begin{equation}
\label{7to3}
\{7\} = \{3\} \oplus \{\overline 3\} \oplus \{1\}.
\end{equation}

Therefore the Cartan generator of the $G(2)$ gauge group using the above decompositions can be constructed by the $SU(3)$ Cartan generators,
\begin{equation}
\label{Cartan 1}
\mathcal{H}_a^{7} = \frac{1}{\sqrt{2}} \left( \begin{array}{ccc} \mathcal{H}_a^3 &0 & 0 \\
0 & \; 0 & 0\\ 0 & 0 & -(\mathcal{H}_a^3)^* \ \end{array}\right),
\end{equation}
where $\mathcal{H}_a^3$ ($a=3,8$) are the $SU(3)$ Cartan generators in the fundamental representation. 

\acknowledgments
\label{acknowledgments}
We are grateful to the research council of the University of Tehran for supporting this study.


\begin{thebibliography}{xx}
%
\bibitem{Fabe1998}
  M.~Faber, J.~Greensite, and S.~Olejnik,
  \emph{Casimir scaling from center vortices: towards
an understanding of the adjoint string tension},
  \emph{Phys. Rev.}  {\bf D 57} (1998) 2603.
 \bibitem{Greensite2002}
  J.~Greensite, C.B.~Thornk,
  \emph{Gluon chain model of the confining force},
  \emph{JHEP}  {\bf 014} (2001) 0202.
\bibitem{Ripka}
  G.~Ripka, \emph{Dual superconductor models of color confinement}, arXiv:hep-ph/0310102v2.
\bibitem{review}
  J.~Greensite,
  \emph{The confinement problem in lattice gauge theory},
  \emph{Prog. Part. Nucl. Phys.}  {\bf 51} (2003) 1.
\bibitem{michael}
  M.~Engelhardt,
  \emph{Generation of confinement and other nonperturbative effects by infrared gluonic degrees of freedom},
  \emph{Nucl. Phys. Proc. Suppl.}  {\bf 140} (2005) 92.
\bibitem{Hoof79}
  G.~'t Hooft,
  \emph{A property of electric and magnetic flux in non Abelian gauge theories},
  \emph{Nucl. Phys.}  {\bf B 153} (1979) 141.
\bibitem{Corn}
  J.~M.~Cornwall,
  \emph{Quark confinement and vortices in massive gauge invariant QCD},
  \emph{Nucl. Phys.}  {\bf B 157} (1979) 392; G.~Mack, V.B.~Petkova,
  \emph{Comparison of lattice gauge theories with gauge groups $Z_2$ and $SU(2)$},
  \emph{Nucl. Phys. (NY)}  {\bf 123} (1979) 442; 
  \emph{Sufficient condition for confinement of static quarks by a vortex condensation mechanism},
  \emph{Nucl. Phys. (NY)}  {\bf 125} (1980) 117; \emph{$Z_2$ monopoles in the standard $SU(2)$ lattice gauge theory model},
  \emph{Z. Physik}  {\bf C 12} (1982) 177; H.B. Nielson, P. Olesen, \emph{A quantum liquid model for the QCD vacuum: gauge and
rotational invariance of domained and quantized homogeneous color fields},
  \emph{Nucl. Phys.}  {\bf B 160} (1979) 380; J. Ambj{\o}rn, P. Olesen, \emph{On the formation of a random color magnetic quantum
liquid in QCD},
  \emph{Nucl. Phys.}  {\bf B 170} (1980) 60; L.G.~Yaffe, \emph{On the formation of a random color magnetic quantum
liquid in QCD}, \emph{Phys. Rev.}  {\bf D 21} (1980) 1574.
\bibitem{Piccioni2005}
  C.~Piccioni,
  \emph{Casimir scaling in $SU(2)$ lattice gauge theory},
  \emph{Phys. Rev.}  {\bf D 73} (2006) 114509.
\bibitem{Deldar1999}
  S.~Deldar,
  \emph{Static $SU(3)$ potentials for sources in various representations},
  \emph{Phys. Rev.}  {\bf D 62} (2000) 034509; \emph{A new lattice measurement for potentials between static $SU(3)$ sources},
  \emph{Eur. Phys. J.}  {\bf C 47} (2006) 163.
\bibitem{Bali2000}
  G.S.~Bali,
  \emph{Casimir scaling of $SU(3)$ static potentials},
  \emph{Phys. Rev.}  {\bf D 62} (2000) 114503.
\bibitem{Greensite2007}
  J.~Greensite, K.~Langfeld, S.~Olejnik, H.~Reinhardt and T.~Tok,
  \emph{Color screening, Casimir scaling, and domain structure in $G(2)$ and $SU(N)$ gauge theories},
  \emph{Phys. Rev.}  {\bf D 75} (2007) 034501.
\bibitem{Holland2003jy}
  K.~Holland, P.~Minkowski, M.~Pepe and U.~J.~Wiese,
  \emph{Exceptional confinement in $G(2)$ gauge theory},
  \emph{Nucl. Phys.}  {\bf B 668} (2003) 207.
\bibitem{Pepe2005sz}
  M.~Pepe,
  \emph{Confinement and the center of the gauge group},
  \emph{Nucl. Phys. Proc. Suppl.}  {\bf 153} (2006) 207.
\bibitem{Pepe2006er}
  M.~Pepe and U.~J.~Wiese,
  \emph{Exceptional deconfinement in $G_2$ gauge theory,},
  \emph{Nucl. Phys.}  {\bf B 768} (2007) 21.
\bibitem{Cossu2006bi}
  G.~Cossu, M.~D'Elia, A.~Di Giacomo, C.~Pi\-ca and B.~Lucini,
  \emph{Dual superconductivity in $G_2$ group},
  \emph{PoS}  {\bf LAT2006} (2006) 063.
  \bibitem{Cossu2007dk}
  G.~Cossu, M.~D'Elia, A.~Di Giacomo, B.~Lucini and C.~Pi\-ca,
  \emph{$G_2$ gauge theory at finite temperature,},
  \emph{JHEP}  {\bf 10} (2007) 100.
  \bibitem{Cossu2007ns}
  G.~Cossu, M.~D'Elia, A.~Di Giacomo, B.~Lucini and C.~Pi\-ca,
  \emph{Confinement: $G_2$ group case},
  \emph{PoS}  {\bf LATTICE2007} (2007) 296.
\bibitem{Maas2007af}
  A.~Maas and S.~Olejnik,
  \emph{A first look at Landau-gauge propagators in $G_2$ Yang-Mills theory},
  \emph{JHEP}  {\bf 02} (2008) 070.
  \bibitem{Olejnik2008}
  L.~Liptak and S.~Olejnik,
  \emph{Casimir scaling in $G_2$ lattice gauge theory},
  \emph{Phys. Rev.}  {\bf D 78} (2008) 074501.
  \bibitem{Deldar2012}
  S.~Deldar, H.~Lookzadeh, and S.~M.~Hosseini Nejad,
  \emph{Confinement in $G(2)$ gauge theories using thick center vortex model and domain structures},
  \emph{Phys. Rev.}  {\bf D 85} (2012) 054501.
 \bibitem{Engelhardt2000}
  M.~Engelhardt, H.~Reinhardt,
  \emph{Center vortex model for the infrared sector of Yang-Mills theory - Confinement and Deconfinement },
  \emph{Nucl.Phys.}  {\bf B585} (2000) 591.
   \bibitem{Rafibakhsh2014}
  S. Rafibakhsh,
  \emph{The effect of $SU(4)$ group factor in the thick center vortex potentials },
  \emph{Phys. Rev.}  {\bf D 89} (2014) 034503. 
  
  \bibitem{Deldar2014}
  S.~M.~Hosseini Nejad and S.~Deldar,
  \emph{Role of the $SU(2)$ and $SU(3)$ subgroups in observing confinement in the $G(2)$ gauge group},
  \emph{Phys. Rev.}  {\bf D 89} (2014) 014510.
   \bibitem{georgi}
  H. Georgi,
  \emph{Lie algebras in particle physics},
  Benjamin/Cummings, Massachusetts, USA 1992.
  \bibitem{Deldar2001}
  S.~Deldar,
  \emph{Potentials between static $SU(3)$ sources in the fat-center-vortices model},
  \emph{JHEP}  {\bf 013} (2001) 0101.

\end{thebibliography}
\end{document}